\documentclass[twocolumn,amsmath,amssymb,aps,floatfix,superscriptaddress]{revtex4-2}

\usepackage{amsfonts}
\usepackage{amssymb}
\usepackage{amsthm,amsmath}
\usepackage{mathrsfs}
\usepackage{indentfirst}
\usepackage{multirow}
\usepackage{graphicx}
\usepackage{amsmath}
\usepackage[amssymb]{SIunits}
\usepackage{amssymb}
\usepackage{color}
\usepackage[usenames,dvipsnames]{xcolor}
\usepackage{natbib}
\usepackage[T1]{fontenc}
\usepackage[hidelinks, bookmarks=true]{hyperref}
\usepackage{bm}
\usepackage{booktabs}
\usepackage{setspace}

\usepackage{graphicx}
\usepackage{dcolumn}
\usepackage{bm}
\usepackage[usenames,dvipsnames]{xcolor}
\usepackage[mathlines]{lineno}%
\usepackage{tikz}

\newcommand{\um}{~\mu \mathrm{m}}

\hypersetup{
	colorlinks = true,
  	linkcolor=blue,
	citecolor=blue,
	urlcolor=magenta
}
\usepackage{cleveref}

\crefname{figure}{Fig.}{Figs.}
\crefformat{figure}{\textbf{Fig}.~#2#1#3}
\crefname{equation}{Eq.}{Eqs.}
\crefformat{equation}{\textbf{Eq}.~(#2#1#3)}
\crefname{table}{Table}{Tables}
\crefformat{table}{\textbf{Table}.~#2#1#3}

\begin{document}

\title{Active phase-space topology unifies depletion and alignment in bacterial flows}

\author{Mingyang Guan}
 \affiliation{Key Laboratory for Mechanics in Fluid Solid Coupling Systems, Institute of Mechanics, Chinese Academy of Sciences, Beijing 100190, China}
\author{Bowen Ling}
 \affiliation{Key Laboratory for Mechanics in Fluid Solid Coupling Systems, Institute of Mechanics, Chinese Academy of Sciences, Beijing 100190, China}
 \affiliation{School of Engineering Science, University of Chinese Academy of Sciences, Beijing 100049, China}
\author{Enhao Liu}
\affiliation{State Key Laboratory of Deep Earth Processes and Resources, Guangzhou Institute of Geochemistry/Guangdong Provincial Key Laboratory of Mineral Physics and Materials, Guangzhou Institute of Geochemistry, Chinese Academy of Sciences, Guangzhou 510640, China}
\affiliation{Guangdong Research Centre for Strategic Metals and Green Utilization, Guangzhou 510640, China}
\affiliation{University of Chinese Academy of Sciences, Beijing 100049, China}
\author{Guoqian Chen}
\affiliation{Department of Mechanics, Peking University, Beijing 100871, China}
\affiliation{Key Laboratory of River Basin Digital Twinning of Ministry of Water Resources, National Observation and Research Station of Coastal Ecological Environments in Macao,  Macau University of Science and Technology, Macao SAR 999078, China}
\author{Zhan Wang}\email{zwang@imech.ac.cn}
 \affiliation{Key Laboratory for Mechanics in Fluid Solid Coupling Systems, Institute of Mechanics, Chinese Academy of Sciences, Beijing 100190, China}
\affiliation{School of Engineering Science, University of Chinese Academy of Sciences, Beijing 100049, China}

\setcounter{figure}{0}
\renewcommand{\thefigure}{\arabic{figure}}

\renewcommand{\figurename}{\textbf{Fig.}}

\begin{abstract}
Transport at small scales is classically understood within an equilibrium framework, where dispersion theory successfully describes shear-enhanced diffusion for passive particles in the continuum limit. However, as most bacteria can move on their own, their motility in flows, inherently out of thermal equilibrium, fundamentally challenges this framework. A minimal, predictive unified theory of bacterial transport in low-Reynolds-number flows remains lacking. Here, from first principles, we develop an analytical hydrodynamic model that enforces consistent no-flux boundary conditions and uses the method of images to characterize the flow--wall coupling. The model quantitatively reproduces measured bacterial distributions and reveals a hydrodynamic locking mechanism accompanied by mean-drift invariance --- an active counterpart to Taylor dispersion. We clarify that shear-induced depletion and alignment are dual manifestations of a single active phase-space topology, ruling out explanations based solely on the local shear magnitude. The theory is validated against microfluidic experiments spanning multiple bacterial species and shear geometries, from one-dimensional to fully three-dimensional flows. Our findings establish a unified phase-space framework for bacterial hydrodynamics, advancing the fundamental understanding of active matter.
\end{abstract}

\maketitle

Since Einstein's seminal work on stochastic motion in quiescent fluids \citep{einsteinUberMolekularkinetischenTheorie1905}, small-scale transport has been largely understood within an equilibrium or near-equilibrium framework. 
Half a century later, Taylor demonstrated that shear flows can dramatically enhance the effective diffusivity of solutes, establishing the classical dispersion theory \citep{taylorDispersionSolubleMatter1953}. 
However, living micro-organisms continuously inject energy and reorient in response to flow, placing their dynamics intrinsically out of thermal equilibrium and beyond the scope of classical theory \citep{laugaHydrodynamicsSwimmingMicroorganisms2009,marcosBacterialRheotaxis2012,bechingerActiveParticlesComplex2016}. 
This raises a central question: what minimal physical ingredients govern bacterial transport in flows?

Shear-induced responses of micro-swimmers have been widely reported \cite{pedleyHydrodynamicPhenomenaSuspensions1992,durhamDisruptionVerticalMotility2009,zottlNonlinearDynamicsMicroswimmer2012,barryShearinducedOrientationalDynamics2015,katuriCrossstreamMigrationActive2018,aransonBacterialActiveMatter2022,kamdarColloidalNatureComplex2022,weiConfinementReducesSurface2025,heinsSelfinducedFloquetMagnons2026,yeoShearinducedLimitBacterial2026}, often interpreted as the result of local velocity gradients that rotate and bias swimmer trajectories. 
Yet, this intuition is incomplete. 
The dynamics are inherently phase-space processes, arising from the coupled evolution of position and orientation. 
In canonical Poiseuille flows, this coupling generates swinging and tumbling trajectories organized by invariants of motion, with related behaviours observed experimentally \cite{zottlNonlinearDynamicsMicroswimmer2012,uppaluriFlowLoadingInduces2012,omoriRheotaxisMigrationUnsteady2022}. 
In vortical flows, curvature and velocity gradients reshape the phase space in qualitatively distinct ways \cite{taramaDeformableMicroswimmerSwirl2014,arguedas-leivaMicroswimmersAxisymmetricVortex2020}. 
A striking paradox then emerges: bacteria are depleted near the low-shear centreline in channel flows \cite{rusconiBacterialTransportSuppressed2014}, yet are expelled from high-shear cores in vortical flows \cite{sokolovRapidExpulsionMicroswimmers2016}. 
Neither local shear magnitude nor noise alone can reconcile these opposing trends, pointing to a missing organizing principle.

Here, we show that this principle is the active phase-space topology. 
We develop a unified theory in which self-propulsion, hydrodynamic reorientation, and no-flux boundary conditions are incorporated consistently from first principles. 
Without invoking ad hoc surface interactions, the model quantitatively captures both near-wall accumulation and bulk depletion, demonstrating that hydrodynamic depletion emerges intrinsically from the underlying Smoluchowski dynamics. 
Within this analytical framework, shear partitions phase space into trapping and escaping regions separated by sharp separatrices that determine bacterial fate. 
Topological transitions of these structures give rise to a hydrodynamic locking mechanism --- an active analogue of Taylor dispersion --- producing a finite drift that can synchronise with the background flow and, in the ellipsoidal limit, becomes independent of swimming speed and noise. 
Experiments across diverse bacterial species and flow geometries, from simple one-dimensional shear to fully three-dimensional (3D) vortical flows, collapse onto universal curves. 
This collapse reveals that depletion and alignment are not distinct phenomena but dual manifestations of a single topological structure, establishing a unified hydrodynamic framework for bacterial transport and suggesting new strategies for control via the design of flow-induced phase-space geometry.

\textbf{Minimal model of bacterial hydrodynamics.} 
We formulate a first-principles, continuum model for a dilute suspension of elongated, self-propelled bacteria in a viscous fluid.
Each bacterium is modelled as a point-like active particle that swims at a constant intrinsic speed $V_s$ along its instantaneous orientation vector $\bm{p}$, with $|\bm{p}|=1$.
In the low-Reynolds-number regime, inertial effects are negligible and the steady flow field $\bm{v}_f(\bm{r})$ is governed by the Stokes equations, which depends only on spatial position $\bm{r}$.
As a consequence, the dynamics of each swimmer is determined by the linear superposition of two contributions: (i) passive advection by the background flow, and (ii) active self-propulsion.
The resulting translational velocity is therefore
\begin{eqnarray}
	\bm{v} (\bm{r}) = \bm{v}_f(\bm{r}) + V_s \bm{p}.
\end{eqnarray}

The evolution of the orientation is governed by Jeffery's equation for an ellipsoidal body of revolution in a viscous flow \citep{jefferyMotionEllipsoidalParticles1922, saintillanInstabilitiesPatternFormation2008, kimMicrohydrodynamicsPrinciplesSelected2013}. 
The local angular velocity is
\begin{eqnarray}
	\bm{\Omega}(\bm{r},\bm{p}) = \frac{1}{2} \bm{\Omega}_f(\bm{r},\bm{p}) + B~ \bm{p} \bm{\times} \left[\bm{E} {\cdot} \bm{p} \right],
    \label{eq:jeffery}
\end{eqnarray}
characterized by two tensorial quantities derived from the velocity gradient $\bm{\nabla} \bm{v}_f$: 
(i) the vorticity vector $\bm{\Omega}_f = \bm{\nabla} \bm{\times} \bm{v}_f$, which describes the rate of rigid-body rotation of the fluid element, 
and (ii) the rate-of-strain tensor $\bm{E} = (\bm{\nabla}\bm{v}_f + \bm{\nabla}\bm{v}_f^{\text{T}})/2$, which describes the rate of shear deformation. 
The Bretherton parameter $B = (\Lambda^2-1)/(\Lambda^2+1)$ quantifies shape anisotropy, where $\Lambda$ is the aspect ratio. 
For elongated, rod-like bacteria ($B \rightarrow 1$), the swimmer aligns strongly with the principal axis of extension. 
For spherical particles ($B = 0$), the particle simply rotates with half the local vorticity.

To account for stochasticity arising from thermal fluctuations and intrinsic biological noise (e.g. fluctuations in flagellar propulsion), we incorporate translational and rotational diffusion. 
The individual-based dynamics is described by the Langevin equations:
\begin{eqnarray}
	\dot{\bm{r}} = \bm{v} + \bm{\xi}_r,\quad
	\dot{\bm{p}} = \bm{\Omega} \bm{\times} \bm{p} + \bm{\xi}_p.
	\label{eq:Langevin}
\end{eqnarray}
The terms $\bm{\xi}_r(t)$ and $\bm{\xi}_p(t)$ are independent Gaussian white noises, with zero mean and second-order correlations
\begin{eqnarray}
    \langle \xi_{r,i}(t)\xi_{r,j}(\tau)\rangle = 2D_t \delta_{ij}\delta(t-\tau),\nonumber\\ 
    \langle \xi_{p,i}(t)\xi_{p,j}(\tau)\rangle = 2D_r \delta_{ij}\delta(t-\tau). 
\end{eqnarray}
Here, $D_t$ and $D_r$ denote the translational and rotational diffusivities, respectively. 
The rotational diffusivity defines a key physical timescale, the persistence time $\tau_r \sim 1/D_r$ over which a swimmer retains memory of its orientation.

At the population level, the bacterial dynamics is described by the probability density $P(\bm{r},\bm{p},t)$ in the combined ${r}$–${p}$ phase space. 
Conservation of probability yields the Smoluchowski equation
\begin{eqnarray}
\partial_t P
+\nabla_{{r}}\cdot \bm{J}_{{r}}
+\nabla_{{p}}\cdot \bm{J}_{{p}}=0,
\label{eq:FPE}
\end{eqnarray}
with probability fluxes defined as
\begin{eqnarray}
\bm{J}_{r}=(\bm{v}_f+V_s\bm{p})P-D_t\nabla_{{r}}P,
\quad \bm{J}_{{p}}=\dot{{p}}\,P-D_r\nabla_{{p}}P.
\end{eqnarray}
The first terms in these fluxes represent deterministic drift (convection and Jeffery orbit alignment), while the second terms capture the diffusive spreading due to random motility. 
This equation provides a complete statistical description linking single-particle dynamics to macroscopic observables such as concentration profiles and orientational distributions.

A central challenge in active matter is the formulation of physically consistent boundary conditions, e.g. antisymmetric periodic \citep{rusconiBacterialTransportSuppressed2014}, specular reflecting \citep{ezhilanTransportDiluteActive2015}, absorbing--reflecting \citep{tanasijevicMicroswimmersVorticesDynamics2022}, and integral conditions \citep{bearonTrappingHighshearRegions2015}. 
In standard Fickian diffusion or dispersion problems, the spatial operator is self-adjoint, and the boundary conditions are typically of the Dirichlet or Neumann type, allowing direct application of Sturm–Liouville theory and separation of variables.
Unlike passive particles, micro-swimmers can interact with boundaries through a combination of hydrodynamic, steric, and propulsion-induced effects, leading to a wide range of phenomenological models and boundary constraints.
Direct comparisons between theoretical predictions and controlled experiments have, however, often revealed systematic discrepancies, particularly regarding the magnitude and spatial extent of near-wall accumulation layers. 
A consensus on theoretical boundary conditions has therefore been elusive \citep{zengSharpTurnsGyrotaxis2022}.

In this Article, we adopt the no-flux condition in position space 
\begin{eqnarray}
	\bm{J}_r {\cdot} \bm{n} =0,
    \label{robin_bc}
\end{eqnarray}
where $\bm{n}$ is the unit outward normal.
This is the natural consequence of impermeability at the macroscopic level \citep{brennerMacrotransportProcesses1993,ezhilanTransportDiluteActive2015,jiangDispersionActiveParticles2019}. 
Unlike passive particles, where zero flux implies a zero concentration gradient, the active flux yields a Robin-type boundary condition.
This coupling between translation and orientation creates an intrinsically non-equilibrium mathematical structure. 
Direct separation of variables fails, and the resulting eigenfunctions are not mutually orthogonal under the standard inner product. 
Analytical solutions of the Smoluchowski equation can be constructed using eigenfunction expansions, see \textbf{Supplementary Note 1}.

The first key theoretical advance lies in analytically resolving this intrinsically non-equilibrium boundary value problem.
By constructing an appropriate transformed operator and a generalized eigenfunction framework (\textbf{Supplementary Note 1}), we show that the derived solutions quantitatively capture both near-wall accumulation and bulk depletion, without invoking any additional surface interactions.
Crucially, this establishes that hydrodynamic depletion is not a consequence of ad hoc boundary physics, but emerges inherently from the interplay between active propulsion, flow-induced reorientation, and the no-flux constraint.

Remarkably, we find that this single boundary condition, when coupled with the full Smoluchowski dynamics, quantitatively agrees with experimental measurements of bacterial locking and depletion across a broad parameter space. 
The validation encompasses five distinct bacterial species with varying sizes, shapes, and swimming speeds, as well as canonical flow configurations spanning one to three spatial dimensions---including planar shear flows and axisymmetric vortical flows. 
The consistent agreement observed across this diverse set of systems strongly suggests that the no-flux condition, despite its simplicity, captures the essential physics governing hydrodynamic depletion at the continuum level. 
It therefore offers a valuable benchmark for future theoretical efforts aimed at resolving the active boundary conditions.

\textbf{Hydrodynamic theory of statistical invariance.} We consider the canonical 3D Stokes-flow problem of a rigid spherical particle of radius $R$ rotating with constant angular velocity $\bm{\omega}$ in an otherwise quiescent, incompressible Newtonian fluid, characterized by its dynamic viscosity $\mu$ and density $\rho$.
At low Reynolds numbers,
$
\mathrm{Re}={\rho\,\omega R^2}/{\mu}\ll 1,
$
the velocity field $\bm{u}(\bm{r})$ and pressure field $p(\bm{r})$ thus satisfy the steady Stokes equations
$
-\nabla p+\mu\nabla^2\bm{u}=0,~
\nabla\cdot\bm{u}=0.
$
The boundary constraints are the no-slip condition on the surface of the rotating particle and the infinite quiescence condition far from the particle:
$
\bm{u}(r=R)=\bm{\omega}\times\bm{r},~
\bm{u}(r\to\infty)=0.
$
Here, $r=|\bm{r}|$ is the distance from the particle centre.
Because the motion is purely rotational, no radial or axial flow is generated.
Exploiting the axial symmetry of the problem, the exact solution can be expressed as a rotlet, i.e. the fundamental singular solution of the Stokes flow:
\begin{eqnarray}
	\bm{u}(\bm{r})
	=\frac{R^3}{r^3}\,(\bm{\omega}\times\bm{r}).
\end{eqnarray}
The velocity magnitude decays algebraically as $|\bm{u}| \sim r^{-2}$, reflecting the long-range nature of viscous disturbances at low Reynolds numbers. 
The flow is purely azimuthal and carries non-zero vorticity, distinguishing it from irrotational potential vortices despite a superficial similarity in streamline geometry. 
Physically, this field represents the diffusion of angular momentum away from a localized torque source.

\begin{figure*}
	\includegraphics[width=0.85\linewidth]{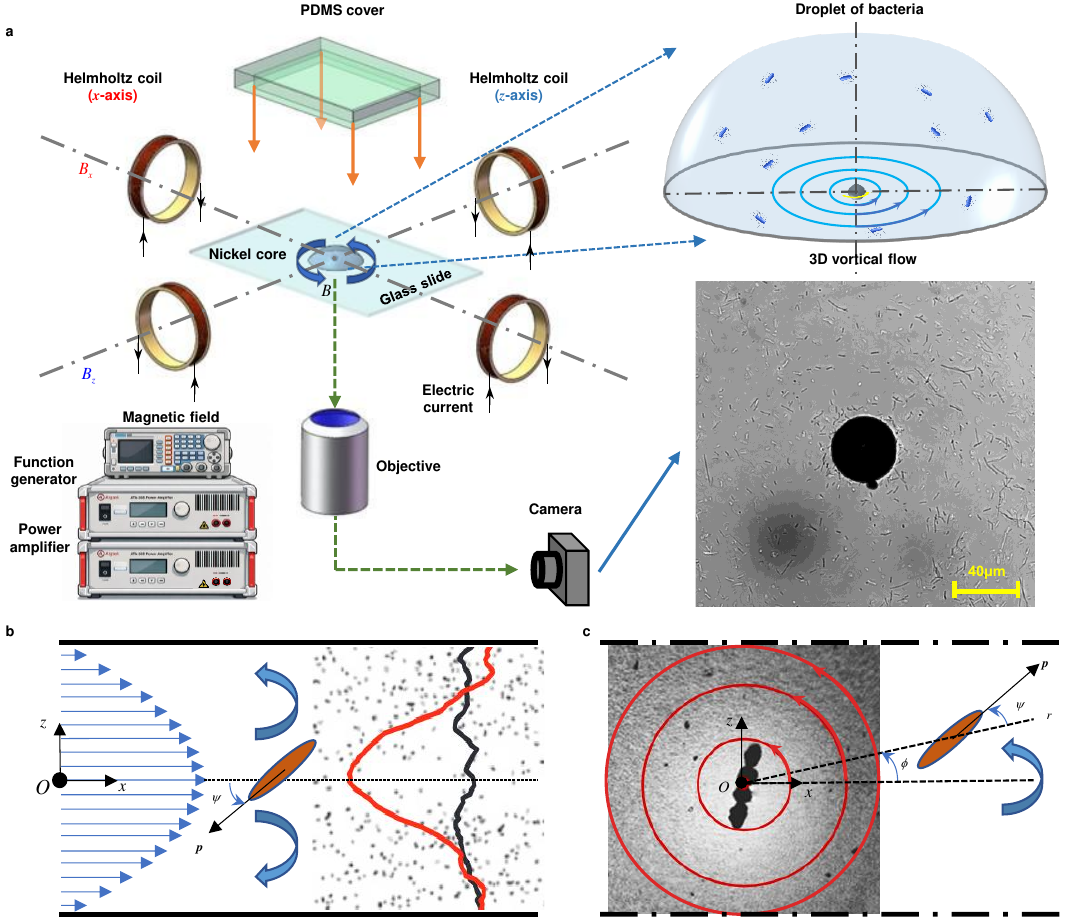}
	\caption{
		\textbf{Experimental setup and shear-induced depletion of bacteria in distinct flows.} 
        \textbf{a}, 3D vortical microfluidic platform.
        A uniform rotating magnetic field $B$, generated by orthogonal Helmholtz coils, drives a nickel particle to spin near a flat plate at the nadir of a sessile bacterial droplet.
        The no-slip boundary breaks axial symmetry and transforms the nominal rotlet into a confined vortex with a high-shear annulus.
        Optical imaging reveals pronounced bacterial depletion near the vortex core with strong boundary adhesion.
        This integrated platform enables controlled magnetic actuation, hydrodynamic confinement, and cell imaging in three dimensions.
		\textbf{b}, Depletion in planar Poiseuille flows occurs in the low-shear region.
		The shear rate (cyan arrow) changes sign across the channel centreline.
        \textit{B.~subtilis} showed depletion near the centerline (red concentration profile), compared with the initial uniform distribution (black).
		Experimental data adapted with permission from Rusconi, Guasto, and Stocker in Ref. \citep[][Fig.~1h]{rusconiBacterialTransportSuppressed2014}.
		\textbf{c}, Depletion in rotlet flows occurs in the high-shear region.
		The central depletion zone formed rapidly upon the onset of rotation.
		Experimental snapshots adapted from Sokolov and Aranson \citep[][Fig.~2b]{sokolovRapidExpulsionMicroswimmers2016} under the creative commons license, 
		\url{http://creativecommons.org/licenses/by/4.0} \textcopyright{} (2016) The Author(s).
	}
	\label{fig1}
\end{figure*}

In experimental realizations shown in \cref{fig1}a, bacterial motion is effectively confined to a thin horizontal layer adjacent to a solid substrate. 
Restricting attention to the mid-plane and assuming the rotation axis is normal to this plane, the 3D rotlet reduces to a quasi-two-dimensional azimuthal flow
$
\bm{u}
=
u_\phi(r)\,\hat{\bm{e}}_\phi,~
u_\phi(r)\simeq {R^3\omega}/{r^2}.
$

\begin{figure}
	\includegraphics[width=0.85\linewidth]{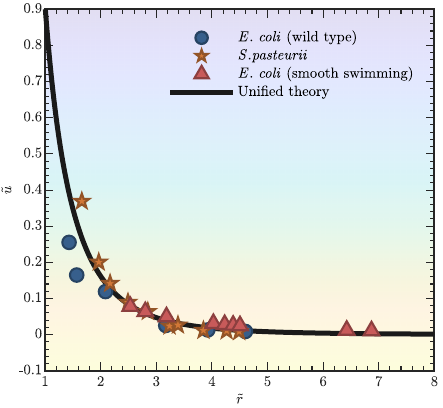}
	\caption{
		\textbf{Hydrodynamic locking mechanism across bacterial species and local positions.} 
        Experimental measurements of local azimuthal drift for \textit{E. coli} (wild type), \textit{S. pasteurii}, and \textit{E. coli} (smooth swimming) are well captured by the unified theory with image corrections (black line).
        The slight discrepancy near the vortex core arises from 3D flow structures and boundary effects not captured by the idealized Stokes flow and the first-order image correction.
	}
	\label{fig2}
\end{figure}

However, the presence of the substrate enforces an additional no-slip boundary condition. 
In the far-field limit ($r \gg R$), the dominant hydrodynamic correction arising from this confinement can be captured by the method of images: a mirror rotlet of equal magnitude and opposite sign is placed symmetrically below the wall to enforce zero velocity on the boundary. 
Introducing dimensionless variables $\tilde{r} = r/R$ and $\tilde{u}_f = u/(R\omega)$, the azimuthal flow velocity takes the form
$\tilde{u}_f(\tilde r)={1}/{\tilde r^2}-{\tilde r}/{(\tilde r^2+4)^{3/2}}$.
The first term represents the direct rotlet flow from the rotating particle, while the second term corresponds to the substrate-induced image singularity. 
Higher-order singularities (e.g. stresslets or source dipoles) become relevant only at distances comparable to $R$. 
The collapse of experimentally measured drift profiles onto a single parameter-free curve (see \cref{fig2}) upon rescaling provides a stringent validation of the hydrodynamic theory.

We now address another central biophysical quantity $\langle \dot{\phi} \rangle(r)$: the mean azimuthal rate of bacterial trajectories orbiting the vortex centre.
From Lagrangian particle tracking, this corresponds to the lab-frame angular drift. 
For a swimmer with instantaneous orientation angle $\psi$ relative to the azimuthal direction, the local angular velocity is given by the superposition of advection by the fluid $\Omega_\phi(r)$ and self-propulsion ${V_s}\sin\psi/{r}$.
The mean azimuthal drift at a given radial position $r$ is obtained by averaging over the steady-state probability distribution of finding a swimmer at radius $r$ with orientation $\psi$ \citep{guanPreasymptoticDispersionActive2023}:
\begin{eqnarray}
	\langle\dot\phi\rangle(r)
	=
	\int_0^{2\pi}
	\left[
	\Omega_\phi(r)
	+
	\frac{V_s}{r}\sin\psi
	\right]
	P(r,\psi)\,\mathrm d\psi.
    \label{eq_azim_drift}
\end{eqnarray}
A key theoretical result emerges when considering the limit of an active prolate spheroid with large aspect ratio  (rigid rod-like bodies such as \textit{E. coli}). In this idealized limit, the mean azimuthal drift becomes exactly locked to the background vortical flow:
$
\langle\dot\phi\rangle(r) = \Omega_\phi(r).
$
Crucially, this locking is independent of the swimming speed $V_s$ and the strengths of thermal or biological noise for the steady state.
This phenomenon arises from a vanishing angular probability flux in \cref{eq_azim_drift} governing the relative orientation $\psi$. 

Specifically, as $B \to 1$, the dominant balance in the steady-state orientation dynamics simplifies dramatically to
$
{\mathrm{d}}/{\mathrm{d}\psi}\!\left[(1+\cos 2\psi)P\right]=0.
$
The solution to this equation, as well as generalized von Mises solution in the limit of weak activity, dictates that the probability density can be regarded as an even function about $\psi$ (orientations aligned with the local flow streamlines).
Since the self-propulsion term involves $\sin\psi$, which is an odd (antisymmetric) function about these directions, the integral over orientation vanishes identically:
$
\int_0^{2\pi}\sin\psi\,P(r,\psi)\,\mathrm d\psi = 0.
$
Physically, activity reshapes the orientation statistics and spatial distributions, but does not modify the mean transport at leading order. 
For finite $B$ or strong noise, only small corrections arise.

The dimensionless mean drift of elongated bacteria reduces to (\textbf{Supplementary Note} 2)
\begin{eqnarray}
	\tilde u(\tilde r) = \dfrac{\langle\dot\phi\rangle r}{\omega R} =\frac{1}{\tilde r^2}
	-\frac{\tilde r}{(\tilde r^2+4)^{3/2}}.
    \label{eq_locking}
\end{eqnarray}
This phenomenon has a direct analogue in classical Taylor dispersion: transverse diffusion enhances dispersion without altering the mean advective velocity. 
Here, orientational dynamics plays the role of an effective transverse stochastic process, leaving the mean drift unchanged.

Biologically, bacteria inhabiting aquatic environments can be efficiently advected by coherent flow structures without actively steering their mean transport direction. 
In nutrient-rich or mixing-dominated habitats, such hydrodynamic locking may represent an evolutionarily advantageous strategy, allowing micro-organisms to sample large volumes of fluid while minimizing the energetic cost of directed navigation against the flow. 
The theory provides a quantitative framework for predicting bacterial transport in microfluidic vortices and marine aggregates, linking microscale flow topology to macroscale bacterial adaptation.

\textbf{Experiments across shear geometries.} In our experiment, a sessile drop of bacterial suspension is attached to a microscope substratem, complementary to those of \cite{sokolovRapidExpulsionMicroswimmers2016}.
A single nickel particle is positioned at the bottom of the drop and held close to the surface by gravity. 
When subjected to a uniform rotating magnetic field generated by two pairs of orthogonal Helmholtz coils, the particle spins steadily and drives a 3D vortical flow (\textbf{Methods}). 
Owing to the proximity of the solid boundary, the resulting flow is not an ideal two-dimensional rotlet but a boundary-modified, 3D vortex with strong shear gradients and broken fore–aft symmetry along the vertical direction. 
This near-wall confinement is essential: it transforms the rotlet flow into a spatially extended vortical structure, with streamlines closing in the horizontal plane while weak vertical recirculation persists.

Experiments consistently show that non-motile bacteria remain uniformly distributed \citep{rusconiBacterialTransportSuppressed2014,sokolovRapidExpulsionMicroswimmers2016}, establishing that depletion is an inherently active phenomenon arising from the coupling between bacterial self-propulsion and fluid shear. Remarkably, expulsion persists even in the absence of noise, indicating a deterministic origin.
Yet, the spatial location of depletion appears contradictory across flow geometries. In planar Poiseuille flow, depletion layers emerge near the channel centreline \citep{rusconiBacterialTransportSuppressed2014}, precisely where the shear rate vanishes and reverses sign (\cref{fig1}b). By contrast, in a rotlet flow, strong depletion occurs near the vortex core \citep{sokolovRapidExpulsionMicroswimmers2016}, where the shear rate is maximal (\cref{fig1}c). These opposing observations rule out explanations based solely on local shear magnitude.

\textbf{Hydrodynamic locking in microfluidics.} As theoretically predicted by \cref{eq_locking} and experimentally measured in \cref{fig2}, the mean drift is robustly locked to the background flow across bacterial species, swimming modes, and vortex strengths (also see \textbf{Supplementary Videos} 1--4).
The physical origin of this locking phenomenon can be traced to the low-Reynolds-number bacterial hydrodynamics; all measurements of $\tilde{u}$ collapse onto a universal curve when expressed in terms of $\tilde{r}$.
By contrast, the unbounded rotlet solution systematically overestimates the drift, underscoring the essential role of confinement.

The mechanism is further corroborated by repeating the experiments in a pendant droplet, in which the influence of the no-slip boundary condition on the substrate could be efficiently eliminated (\textbf{Supplementary Video} 5).
Minor deviations observed near the vortex core can be attributed to 3D flow asymmetries arising from the slight non-sphericity of the rotating nickel particle, as well as higher-order wall effects beyond the leading-order image approximation.

As flow strength increases, the influence of self-propulsion and rotational diffusion is progressively suppressed, approaching limiting cases of vanishing noise or passive tracer dynamics (\textbf{Extended Data} \cref{EDFig1}). 
In planar Poiseuille flows, stronger shear sharpens depletion layers and enhances rheotactic alignment, whereas in vortical flows it drives pronounced boundary accumulation and alignment with the swirling flow (\textbf{Extended Data} \cref{EDFig2}). 
Depletion layers consistently form near the saddle points, highlighting that their qualitative existence is dictated by phase-space topology, while noise and confinement set their precise location (\textbf{Extended Data} \cref{EDFig3}). 

\begin{figure*}
	\includegraphics[width=0.85\linewidth]{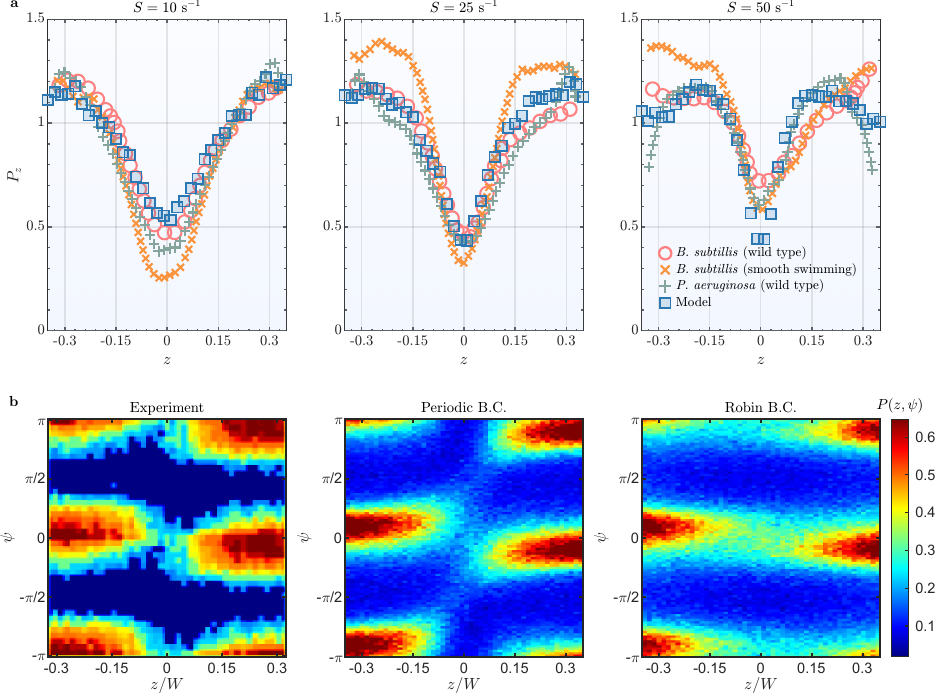}
	\caption{
		\textbf{Steady-state bacterial distributions in unidirectional Poiseuille flows: model versus experiment.} 
		\textbf{a}, Steady-state concentration profiles $P_z$ across Poiseuille channels predicted by the minimal model (blue squares) compared with experimental measurements for wild-type \textit{B.~subtilis}, smooth-swimming \textit{B.~subtilis}, and wild-type \textit{P.~aeruginosa} from Ref.~\citep{rusconiBacterialTransportSuppressed2014}.
		The centreline depletion is robust across species, with quantitative variations attributable to differences in swimming speed and shape (model parameters calibrated to wild-type \textit{B.~subtilis}).
		\textbf{b}, Joint probability distributions $P(z,~\psi)$ in phase space $(z,~\psi)$ from the experiment (left) in Ref.~\citep[][FIG.~3c]{rusconiBacterialTransportSuppressed2014} and the present model with anti-symmetric periodic (middle) and Robin boundary conditions (right). 
	}
	\label{fig3}
\end{figure*}

We now show that, in addition to precise predictions for hydrodynamic locking, our unified theory also applies to bulk depletion and preferential alignment across flow geometries and bacterial species, resolving bacterial hydrodynamics different from that of equilibrium systems.

\textbf{Dynamical origin of depletion and alignment.}
Centreline depletion in Poiseuille flows is a generic outcome of active transport in shear \citep{rusconiBacterialTransportSuppressed2014}, with its magnitude and shear-rate dependence reproduced across diverse bacterial species by our minimal model (\cref{fig3}a). 
Increasing shear transforms depletion into a symmetric double-peaked profile through compression of phase-space separatrices (\textbf{Supplementary} FIG.\ S2). 
Depletion is invariably accompanied by strong rheotactic alignment (upstream-oriented states dominate), and both arise from the same phase-space structures (\cref{fig3}b).

\begin{figure*}
	\includegraphics[width=0.85\linewidth]{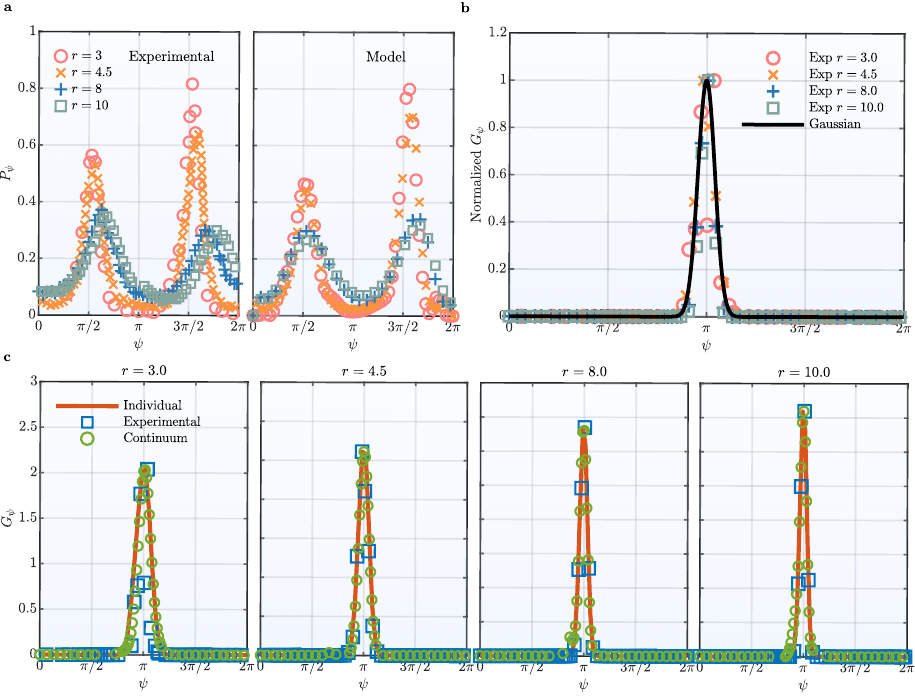}
	\caption{
		\textbf{Steady-state bacterial distributions in 2D vortex flows: model versus experiment.} 
		\textbf{a}, Radius-dependent orientation distributions $P_\psi$ at different radii measured experimentally \citep[left,][]{sokolovRapidExpulsionMicroswimmers2016} and predicted by the present model (right).
		\textbf{b}, Rescaled orientation distributions collapse onto a universal Gaussian via \cref{eq:sol}.
		\textbf{c}, Full orientations $G_\psi$ over $0\le \psi<2\pi$ show quantitative agreement between theory and experiment across radii.
	}
	\label{fig4}
\end{figure*}

In rotlet flows, bacteria align near $\psi=\pi/2$, $\psi=3\pi/2$, offset from exact streamline-following (\cref{fig4}a). 
The strong-vortex limit yields a steady-state distribution
$
P={\sqrt{1-B^2}}/ \big[{2 \pi \left(1+B \cos 2\psi\right)}\big],~ B<1,
$
whose extrema are set by a shape-induced orientational drift $(1+B\cos2\psi)$ (\textbf{Supplementary Note} 3); spherical swimmers ($B=0$) remain isotropic, while rod-like swimmers ($B\to1$) lock onto stagnation points of the deterministic dynamics. 
This alignment emerges from the coupled position–orientation phase-space structure, suppressing cross-stream motion and expelling swimmers from the vortex core.

Further, we exploit the exact solution of \cref{eq:FPE} in the absence of vortex flow and rotational diffusion,
\begin{eqnarray}
	P_a=\exp\left(\frac{{V_s}r \cos\psi}{D_t}\right),
	\label{eq:sol}
\end{eqnarray}
which gives a benchmark distribution under no-flux boundary conditions. 
Normalizing the experimental data by $P_a$ collapses the orientation distributions onto a single Gaussian master curve, independent of radius (\cref{fig4}b,c), revealing a modified diffusive scaling.

\begin{figure}
	\includegraphics[width=0.85\linewidth]{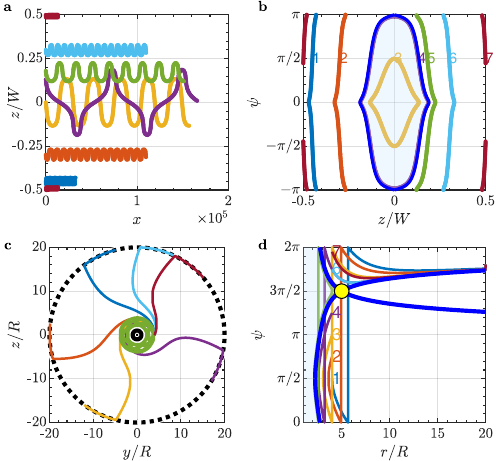}
	\caption{
		\textbf{Separatrices organize bacterial fate: trapping versus escaping.}
        These universal sorting boundaries render depletion and alignment as two facets of the same dynamics. 
		\textbf{a}, Deterministic trajectories in a planar Poiseuille flow.
		\textbf{b}, Phase portrait in the $z-\psi$ plane showing the separatrix (equation (\ref{eq:H_poiseuille}), blue curve) separating tumbling  (escaping) and swinging (trapping) motions.
		Tumbling trajectories ($1$, $2$, and $5$--$7$) are symmetric about $z=0$ and yield no net transverse drift, while swinging orbits (purple) contribute to migration only near the separatrix.
		\textbf{c}, Trapping and escaping trajectories in a rotlet flow.
		The vortex core is surrounded by a family of closed periodic orbits, forming a rosette-like trapping structure (green).
		\textbf{d}, Phase portrait in the $r-\psi$ plane.
		The left branch of the invariant curve (equation (\ref{eq:H_rotlet}), blue curve) defines a separatrix, while the right branch corresponds to asymptotic rheotactic alignment $\psi=3\pi/2$ for escaping swimmers.}
	\label{fig5}
\end{figure}

We further show that both shear-induced depletion and rheotaxis arise from the same active phase-space topology. 
In this view, hydrodynamic shear acts as a deterministic organizer of swimmer trajectories, shaping transport and accumulation independently of biological sensing (\textbf{Supplementary Note} 4). 

\textbf{The active phase-space topology.} In the deterministic limit $(D_t=D_r=0)$, the phase space is organized by separatrices that partition trajectories into qualitatively distinct classes --- trapped orbits that remain confined, and escaping orbits that drift away.
This geometric structure provides a unified, quantitative explanation for two seemingly unrelated phenomena: the formation of depletion layers and rheotactic alignment in shear flows.
Crucially, the underlying mechanism is purely deterministic and kinematic; steric or wall effects are not necessary.

We consider a planar Poiseuille flow between two parallel plates separated by a distance $W$.
The background flow velocity is directed along $x$-axis, $v_f (z)= U (1-4 z^2/W^2)$ where $U$ is the centreline speed, and $z \in [-W/2,~W/2]$ is the transverse coordinate. 
The mean shear rate is $S = 2 U/W$. 
The orientation of the swimmer is described by the angle $\psi$ between its swimming direction and the negative $x$-axis (\cref{fig1}b).

The coupled motion admits a conserved quantity along any trajectory \citep{zottlPeriodicQuasiperiodicMotion2013}:
$	h_1(z,\psi)
	= {z^2}/{2}
	+ K \operatorname{arctanh} \bigl(s\cos\psi\bigr),$
where
$K={V_s W^2}/\big[{4U\sqrt{2B(1+B)}}\big]$ measures the relative importance of swimming to the imposed shear, with a large $K$ (weak flow or fast swimming) expanding the trapping region, and the shape factor $s=\sqrt{{2B}/({1+B})}$ increases monotonically with $B$, reflecting the enhanced alignment of more elongated particles with the flow.
$h_1$ acts as an effective Hamiltonian; invariant curves $h_1=\text{const}$ are the phase separatrix in the $(z,\psi)$ plane, analogous to the phase portrait of a nonlinear pendulum \citep{zottlNonlinearDynamicsMicroswimmer2012}.

The fixed points of equations of motion are found by setting $\dot{z}=\dot{\psi}=0$.
For prolate swimmers, the centreline orientations $\psi=0,\pi$ at $z=0$ are saddle points.
The level set passing through a saddle, $h_1(z,\psi)=h_1(0,0)=K \operatorname{arctanh}(s)$, defines the separatrix, an eye-shaped curve that partitions the phase space (\cref{fig5}b), as 
\begin{equation}
	\frac{z^2}{2}
	= K\Big[\operatorname{arctanh}(s\cos\psi)-\operatorname{arctanh}(s)\Big].
	\label{eq:H_poiseuille}
\end{equation}

Inside the separatrix (\cref{fig5}, shaded blue) trajectories are closed, periodic orbits.
The swimmer oscillates about the centreline with $\phi\simeq0$ and performs symmetric swinging motions; because the orbits are symmetric under $z\to -z$, the vertical displacements cancel over each cycle, yielding no net cross-stream migration.
Trajectories very close to the separatrix, however, break this symmetry, generating a small effective transverse drift (e.g. purple line in \cref{fig5}a). 
These marginal orbits regulate the boundary of the central depletion layer: its width is therefore universally set by the separatrix width \citep{rusconiBacterialTransportSuppressed2014}.

Outside the separatrix, the swimmer enters the high‑shear region where $\psi$ becomes a fast variable.
Phase-space trajectories collapse into nearly vertical strips, corresponding to rapid tumbling.
The fast orientational dynamics average out the vertical motion over a rotation period, preventing crossing of the mid‑plane $z=0$. 
This suppression of cross-stream transport is not of steric or wall origin, but a purely deterministic consequence.
Thus, strong shear alone can exclude swimmers from certain regions of the flow, even without boundary constraints.

For a 2D rotlet flow in polar coordinates $(r,\phi)$, the azimuthal velocity is $v_f (r)= \omega R^2/r$, where $\omega$ sets the vortex strength and $R$ is the vortex core radius.
The swimmer’s orientation $\psi$ is now measured relative to the local radial direction (\cref{fig1}c).
For elongated swimmers, Tanasijevic \& Lauga \citep{tanasijevicMicroswimmersVorticesDynamics2022} derived an invariant: 
$	h_2(r,\psi)
	=\big[{\gamma K_0(\eta)+\sin\psi K_1(\eta)}\big]/\big[
	{\gamma I_0(\eta)-\sin\psi I_1(\eta)}\big]$,
with $\eta=\sqrt{2B(1+B)}/ (V_s r)$ and
$\gamma=\sqrt{(1+B)/(2B)}$.
Here, $I_n$ and $K_n$ are modified Bessel functions of the first and second kind, respectively.
A swimmer is permanently captured by a two-dimensional rotlet if and only if its initial condition ($r_0,\psi_0$) satisfies
\begin{equation}
	r_0< r_c,\quad
	~h_2(r_0,\psi_0)<h_c,
	\label{eq:H_rotlet}
\end{equation}
where $r_c={(1-B)}/{V_s}$ and $h_c={(\gamma K_0- K_1)}/
{(\gamma I_0+ I_1)}$ is the minimum of $h_2$ evaluated at $\eta=\eta(r_c)$.
Inside the separatrix, trajectories form rosette‑like orbits bound to the vortex core; outside, they escape to large radii (\cref{fig5}c,d).

Despite dramatic differences in flow geometries, the dynamics can be cast into a Hamiltonian-like form,
\begin{eqnarray}
	\dot q = \partial_p H(q,p),\qquad
	\dot p = -\partial_q H(q,p),
\end{eqnarray}
with the conserved quantity $H(q,p)$ that acts as a streamfunction in phase plane (\textbf{Supplementary Note} 5).
For planar Poiseuille flows $(q,p)=(z,\psi)$, while for the rotlets $(q,p)=(r,\psi)$.
In each case, the level set $H=H_{\text{saddle}}$ defines a separatrix that divides phase space into trapped and escaping orbits.
This phase‑space topology reveals that shear‑induced depletion and alignment are not stochastic or boundary‑driven phenomena, but rather emerge from the deterministic, nonlinear coupling between orientation and position.
Across shear geometries, phase-space topology emerges as an organizing principle of bacterial dynamics, delineating trajectories that contribute to transport from those that are dynamically suppressed. 

Taken together, our results uncover the striking richness of bacterial dynamics under confinement.
The unified theory resolves a long-standing contradiction: the apparent inversion, i.e. depletion near low-shear centrelines in Poiseuille flows and from high-shear cores in vortices, reflects not competing physical mechanisms, but distinct topological organizations of the same underlying dynamics.
Transport is therefore governed by phase-space topology, not by local shear magnitude.
Supported by experimental evidence, our framework establishes a tractable analytical route to non-equilibrium transport in bacterial hydrodynamics, grounded in physically consistent boundary conditions.

\textbf{Discussion and conclusions.}
By integrating microfluidic experiments, first-principles hydrodynamics, and active phase-space topolgy, we establish a unified theory of bacterial hydrodynamics in confined flows across one, two, and three dimensions. 
Depletion and alignment of motile bacteria emerge not as separate phenomena, but as dual consequences of a single mechanism, divided by sharp separatrices that select swimmer trajectories. 
What the field had treated as separate effects is indeed one unified principle---the kind of surprising, counterintuitive insight that reframes how one thinks about low-Reynolds-number living fluids.

Remarkably, a minimal hydrodynamic model, treating bacteria as point-like elongated swimmers in Stokes flow with no-flux boundaries, quantitatively captures all observations without fitting parameters or behavioural assumptions.
Noise and confinement modulate how trajectories are sampled but leave the deterministic backbone intact. 
This identifies a robust and experimentally testable control principle: bacterial transport can be directed by reshaping phase-space topology through flow design.
Our unified theory highlights the basic coupling between hydrodynamics and bacterial motility, providing a mechanistic perspective for the organization of active matter.
More broadly, it places shear-induced active transport on a unified footing relevant to microfluidics \citep{aminianHowBoundariesShape2016,ozcelikAcousticTweezersLife2018}, precision medicine \citep{el-sayedibrahimLivingMicrorobotsTarget2023,liPreciseElectrokineticPosition2023}, targeted delivery \citep{dasguptaNanoparticleDeliveryTumours2024,leverge-serandourActiveFluidsNavigate2024}, particle sorting \citep{caoOrientationalDirectionalLocking2019,wangTopologicalWaterwaveStructures2025},
and bio-inspired manipulation of active matter \citep{stockerMarineMicrobesSee2012,theryControllingConfinedCollective2024}.

\clearpage
\newpage
\section*{Methods}\label{sec11}
\textbf{Bacterial culture.} \textit{Escherichia coli}, \textit{Sporosarcina pasteurii}, \textit{Bacillus subtilis}, and \textit{P.~aeruginosa} were employed as model micro-organisms. 
Bacteria were streaked on LB agar plates and incubated at 35~$^\circ$C.
Twelve hours prior to each experiment, cells from a single isolated colony were inoculated into 15~mL of liquid LB medium and grown in a shaking incubator at 35~$^\circ$C.
Before use, bacteria were harvested by centrifugation, washed, and resuspended in fresh medium.

Bacterial motility was systematically tuned by adjusting culture conditions, including medium composition, temperature, and nutrient availability.
Swimming behaviour in quiescent fluid was characterized using optical microscopy combined with particle tracking, from which we extracted the effective transport parameters used as inputs to the continuum model.

\textbf{Microfluidic apparatus.} To generate confined vortical flows with precisely tunable parameters, we developed a magnetically actuated microfluidic platform that enables stable and continuous control of vortex structures.
The system comprises a magnetic actuation module, a confinement chamber, and an optical imaging unit.
The actuation module consists of two orthogonally arranged pairs of Helmholtz coils, producing a spatially uniform rotating magnetic field within the chamber (\cref{fig1}c).

The confinement chamber is formed by a glass slide and a polydimethylsiloxane (PDMS) cover.
A nickel particle of radius $\approx 20\mu\mathrm{m}$ was placed on the glass surface, followed by deposition of a 10~$\mu$L droplet of bacterial suspension, which was subsequently sealed with the PDMS cover to suppress evaporation.
Under gravity, the nickel particle remained close to the glass surface throughout the experiment.
When subjected to the external rotating magnetic field, the magnetic particle spun steadily, generating three-dimensional vortical flow near the no-slip boundary.

\textbf{Imaging and tracking.} Bacterial motion was recorded in real time using a high-resolution optical microscope coupled with a high-speed camera. 
Phase-contrast imaging was employed to enhance the signal-to-noise ratio of bacterial contours, improving the accuracy of trajectory extraction.
Trajectory data were processed in ImageJ with background subtraction, noise filtering, and intensity thresholding to improve bacterial detection. 
Automated tracking was performed using the TrackMate plugin, yielding time-resolved positional and orientational trajectories of individual bacteria.
The reconstructed trajectories define the translational and orientational degrees of freedom of active particles, and form the basis for subsequent statistical analysis of bacterial transport in vortex-dominated flows.

\textbf{Analytical solutions for Robin conditions.}
We describe the dynamics of bacteria by the probability distribution $P(\bm{r},\bm{p},t)$ through the Smoluchowski equation
$
\partial_t P
+\nabla_{\bm{r}}\cdot \bm{J}_{\bm{r}}
+\nabla_{\bm{p}}\cdot \bm{J}_{\bm{p}}=0,
$
with translational and orientational probability fluxes
$
\bm{J}_{\bm{r}}=(\bm{v}_f+V_s\bm{p})P-D_t\nabla_{\bm{r}}P,
~\bm{J}_{\bm{p}}=\dot{\bm{p}}\,P-D_r\nabla_{\bm{p}}P.
$
For the migration of bacteria in vortical flows, we can define a local operator
$
\mathcal{L}=
{V_s\cos\psi}{\partial_r(rP)}/{r}
-
{D_t}{\partial_r}
\left(r{\partial_r P}\right) /{r}
-{D_t}{\partial_{\psi\!\psi}^2 P}/{r^2}
-{\partial_\psi}
\left[
{V_s}\sin\psi\,P/{r}
-{\Omega_f(r)}(1+B\cos2\psi)P/2
\right]
-D_r{\partial_{\psi\!\psi}^2 P},
$
so that 
\begin{eqnarray}
    \mathcal{L}P=0,
\end{eqnarray}
for the steady state.
No-flux boundary conditions are imposed at the inner and outer radii,
$
V_s\cos\psi\,P - D_t{\partial_r P}=0,
$
while periodicity holds in the angular coordinate,
$
P(r,0,t)=P(r,2\pi,t),~
\partial_\psi P(r,0,t)=\partial_\psi P(r,2\pi,t).
$
To facilitate analytical treatment of the Robin boundary condition, we factorize the probability distribution as
$
P(r,\psi,t)=P_a(r,\psi)\,G(r,\psi,t),
$
which rigorously transforms the Robin condition into homogeneous Neumann conditions,
$
\partial_r G=0~\text{at}~ r=1,\;r=r_m.
$
The transformed distribution $G$ satisfies
\begin{eqnarray}
\partial_t G
+\mathcal{L}_0 G = 0,
\label{eq:transient}
\end{eqnarray}
with $\mathcal{L}_0(\cdot)=\mathcal{L}\big[P_a\,(\cdot)\big]/P_a$. Analytical solutions are available through eigenfunction expansions (see \textbf{Supplementary Note} 1 for details).

\textbf{Numerical algorithms.} Equation (\ref{eq:Langevin}) was integrated numerically for $10^7$ cells using the Euler–Maruyama scheme. 
Robin-type boundary conditions were imposed at both the vortex core and the droplet boundary, with periodicity in the angular coordinate. 
We verified numerical convergence by systematically reducing the dimensionless time step normalized by the characteristic rotational diffusion time; a value of $\Delta t = 10^{-4}$ was found to be sufficiently small, with further reductions producing no measurable change. 
All model parameters governing cell dynamics, including swimming speed, cell elongation, and rotational noise, were determined independently from experiments or calculated under conditions matching those of the experiments.

The dynamics in vortical flows was modelled by $\bm{R} =(r, \phi)$ and $\bm{p}= (\cos(\psi+\phi),\sin(\psi+\phi))$.
The governing stochastic equations are
\begin{equation}
	\left.
	\begin{aligned}
		&\dot{r} = V_s \cos\psi + \xi_t, \\
		&\dot{\phi} = \dfrac{V_s}{r}\sin\psi + \dfrac{\omega}{r^{ 2}} + \xi_t, \\
		&\dot{\psi} = -\dfrac{\omega}{r^{2}}\!\left(1+B \cos 2\psi\right) - \dfrac{V_s}{r}\sin\psi + \xi_{r}.
	\end{aligned}
	\right\}
\end{equation}
In Cartesian coordinates ($y,~z$), the equivalent individual-based model reads
\begin{equation}
	\left.
	\begin{aligned}
		&\mathrm{d} y = V_s \cos(\phi+\psi) \mathrm{d} t +\sqrt{2D_t} dW_{t},\\
		&\mathrm{d} z = V_s \sin(\phi+\psi) \mathrm{d} t +\sqrt{2D_t} dW_{t},\\
		&\mathrm{d} \psi = -\dfrac{V_s}{r}\sin\psi \mathrm{d} t-
		\dfrac{\omega}{r^2}\left(1+B \cos2\psi\right)\mathrm{d} t+\sqrt{2} \mathrm{d} W_{r},
	\end{aligned}\right\}
\end{equation}
with independent Wiener processes $\mathrm{d}W_t$ and $\mathrm{d}W_r$, and diffusivities $D_t$ (translational) and unity (rotational).
Rotational symmetry ensures confinement of motion to the $y$--$z$ plane.

Confinement at the inner and outer radii ($r=1$ and $r=r_m$) was enforced by no-flux (Robin-type) boundary conditions. In the individual-based model, swimmers attempting to cross the boundary were radially reset to $r=1$ or $r=r_m$, with orientations reoriented according to
$\bm{p} \to \bm{p} - \beta (\bm{p} \cdot \bm{e}_r) \bm{e}_r,
$
$\bm{e}_r = {\bm{r}}/{|\bm{r}|}$
The parameter $\beta$ interpolates between specular reflection ($\beta = 2$), tangential reorientation ($\beta = 1$) and Robin no flux conditions ($\beta = 0$).
The orientation vector was renormalised after each update.

The transformed transient Smoluchowski equation (\ref{eq:transient}) was also solved numerically for validation using a hybrid spectral–finite-difference scheme in polar coordinates.
The radial domain $r\in[1,r_m]$ was discretized on a uniform grid with $N_r$ collocation points, while the angular coordinate $\psi\in[0,2\pi)$ was represented on a periodic grid with $N_\psi$ modes. 
Angular derivatives were evaluated spectrally using Fourier transforms, whereas radial derivatives were computed by second-order finite differences. 
Homogeneous Neumann boundary conditions in the radial direction were enforced by mirrored ghost points, ensuring vanishing normal gradients at both boundaries.
The transformed equation was advanced in time using an explicit Euler scheme. Convective terms in the radial direction were discretized by centred finite differences, and angular advection and diffusion terms were evaluated in Fourier space. 
At each time step, the probability distribution was reconstructed as $P=P_a G$ and renormalized to enforce global probability conservation,
$
\int_0^{2\pi}\!\!\int_{1}^{r_m} P(r,\psi,t)\,r\,dr\,d\psi = 1.
$
The renormalized distribution was then projected back onto $G$ for the subsequent time step.
Simulations were performed with sufficiently small time steps and spatial resolutions to ensure numerical stability and convergence. 
Parameters used in numerical computations and experiments are listed in \textbf{Supplementary Table} 1.
Steady-state solutions were obtained by long-time integration, and marginal distributions in the radial and angular directions were computed by numerical quadrature.
The two numerical methods are in excellent agreement with the analytical solutions and with each other (see \textbf{Extended Data Figs.} \ref{EDFig4} to \ref{EDFig6}).

\clearpage
\newpage
\bibliography{mylib}

\section*{Acknowledgments}
We are grateful to Roberto Rusconi and Roman Stocker for sharing experimental data through private communications. 
We thank Demetrios T. Papageorgiou for valuable insights and broader implications for Stokes-flow transport and low-Reynolds-number hydrodynamics.
We thank Li Zeng, Weiquan Jiang, Bohan Wang, and Hanhan Zeng for insightful discussions and comments on the manuscript. 
We also acknowledge Mian Long, Dongshi Guan, Xu Zheng, Ning Li, and Rongliang Xu for their laboratory assistance on cell culture and technical advice on instrumentation.
M.G. discloses support for the research of this work from the National Natural Science Foundation of China [grant number 123B2039], the Fellowship Program of CPSF [grant number GZC20251266], and Young Scientists Fund of Key Laboratory for Mechanics in Fluid Solid Coupling Systems, Chinese Academy of Sciences. 
Z.W. discloses support from the National Natural Science Foundation of China [grant numbers 12325207, 42450111, and 12132018], and the International Partnership Program of the Chinese Academy of Sciences [grant number 025GJHZ2024047GC].

\section*{\label{sec:contributions}Author contributions}
M.G. and Z.W. designed the research. Z.W. supervised the project. M.G., B.L., and E.L. performed experiments and analyzed data. M.G., Z.W., G.C., and D.T.P. developed analytical tools and performed simulations. All the authors wrote the paper.

\section*{Competing interests}
The authors declare no competing interests.

\section*{Additional information}
\textbf{Supplementary information}: The online version contains supplementary material available at [url to be inserted].
\textbf{Correspondence and requests for materials} should be addressed to Z.W.

\newpage
\setcounter{figure}{0}
\renewcommand{\thefigure}{\arabic{figure}}

\renewcommand{\figurename}{\textbf{Extended Data Fig.}}

\onecolumngrid

\newpage
\begin{figure*}
	\includegraphics[width=0.9\linewidth]{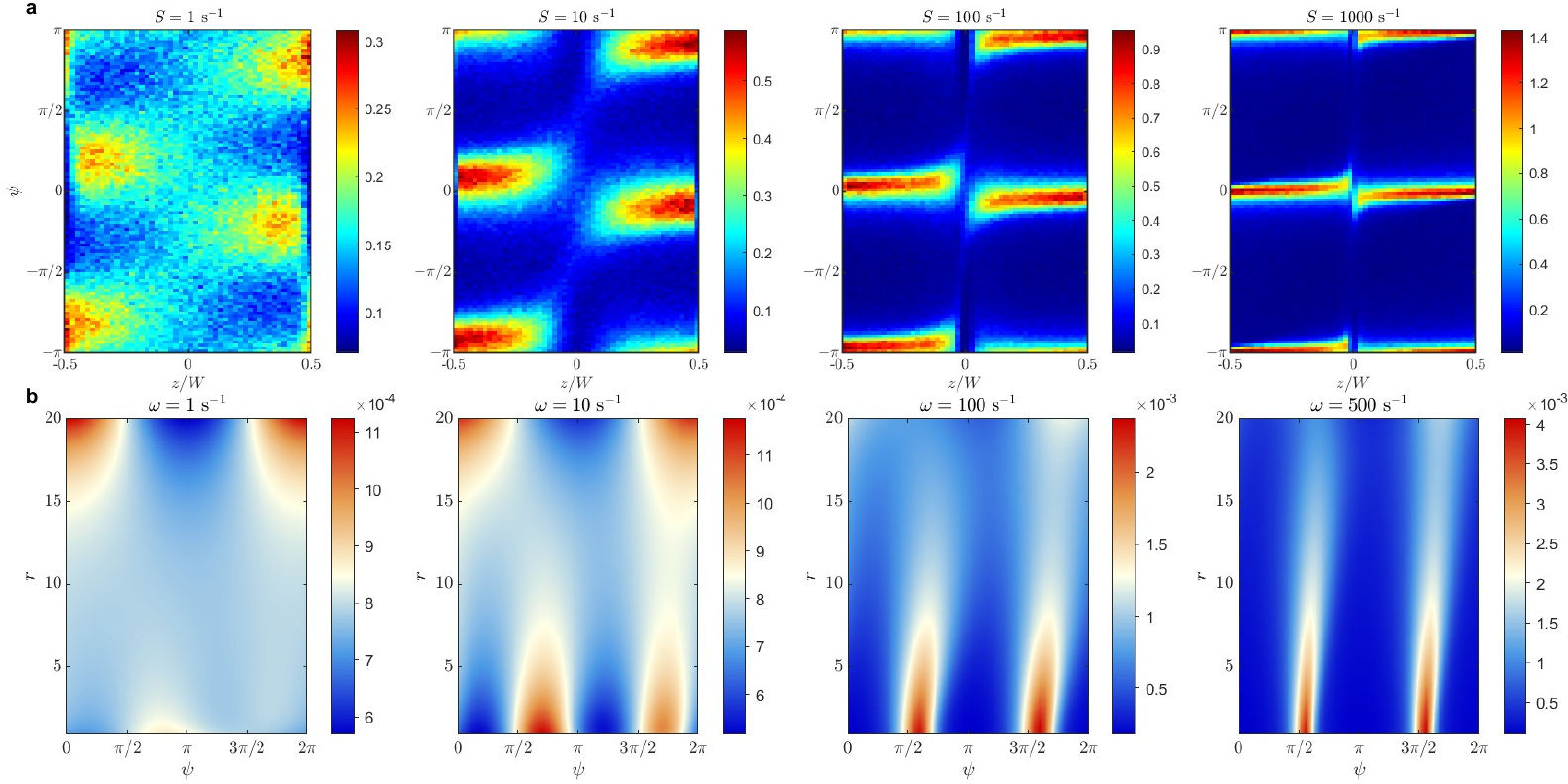}
	\caption{
		\textbf{Unified picture of shear-induced bacterial depletion and rheotaxis.}
		\textbf{a}, In planar Poiseuille flows, fluid shear depletes bacteria from the centreline, producing antisymmetric concentration peaks whose separation is set by the separatrix width.
		\textbf{b}, In rotlet flows, vortical shear expels bacteria from the vortex core, while trapped trajectories collapse toward the centre, and escaping cells migrate radially outward.
	}
	\label{EDFig1}
\end{figure*}

\clearpage
\newpage
\begin{figure*}[ht!]
	\centering
	\includegraphics{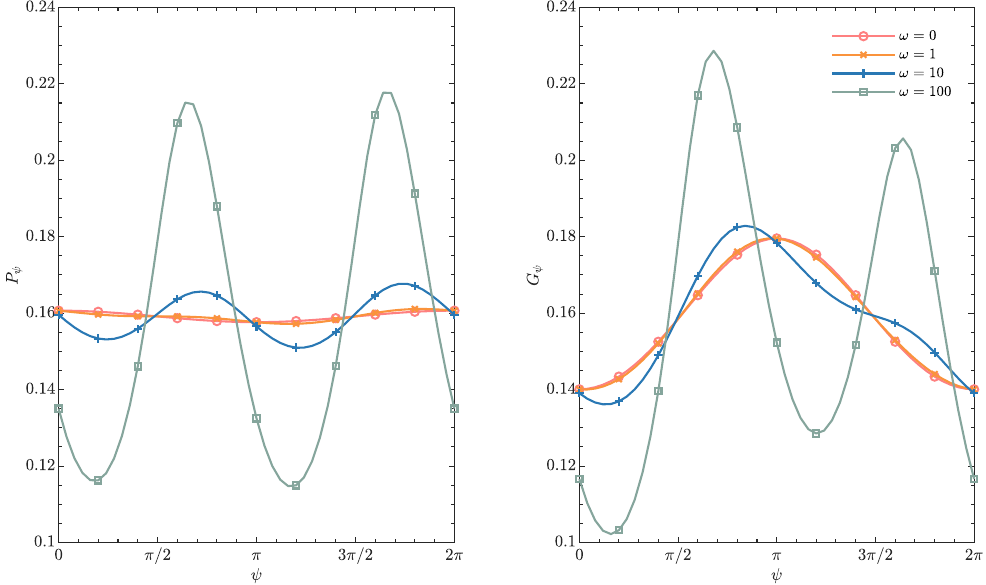}
	\caption{
		\textbf{Effect of translational diffusivity on steady-state bacterial concentration.}
		Decrease in $D_t$ leads to centripetal focusing of $P_r$ in the vicinity of vortex core, as a result of the Robin boundary condition.
	}
	\label{EDFig2}
\end{figure*}
\newpage
\begin{figure*}[ht!]
	\centering
	\includegraphics[width=0.75\linewidth]{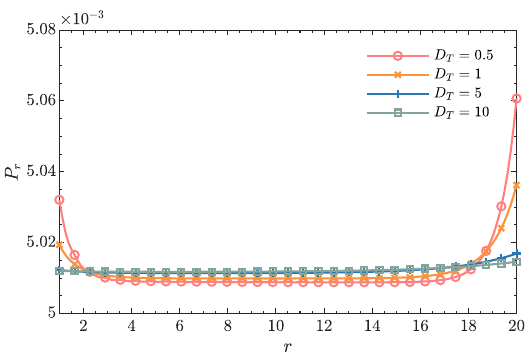}
	\caption{
		\textbf{ Steady-state orientational distributions of bacteria for different vortex strengths.}
		The Gaussian mode of $G_\psi$ observed in the experiment is reproduced robustly within different parameter spaces.
		$G_\psi$ can exhibit various modes for the steady state, and therefore the quantitative agreement shown in the main text is remarkable.
		Besides the Gaussianity, skewed and bimodal modes are also reported for stronger vortices.
	}
	\label{EDFig3}
\end{figure*}

\newpage

\begin{figure*}[ht!]
	\centering
	\includegraphics[width=0.9\linewidth]{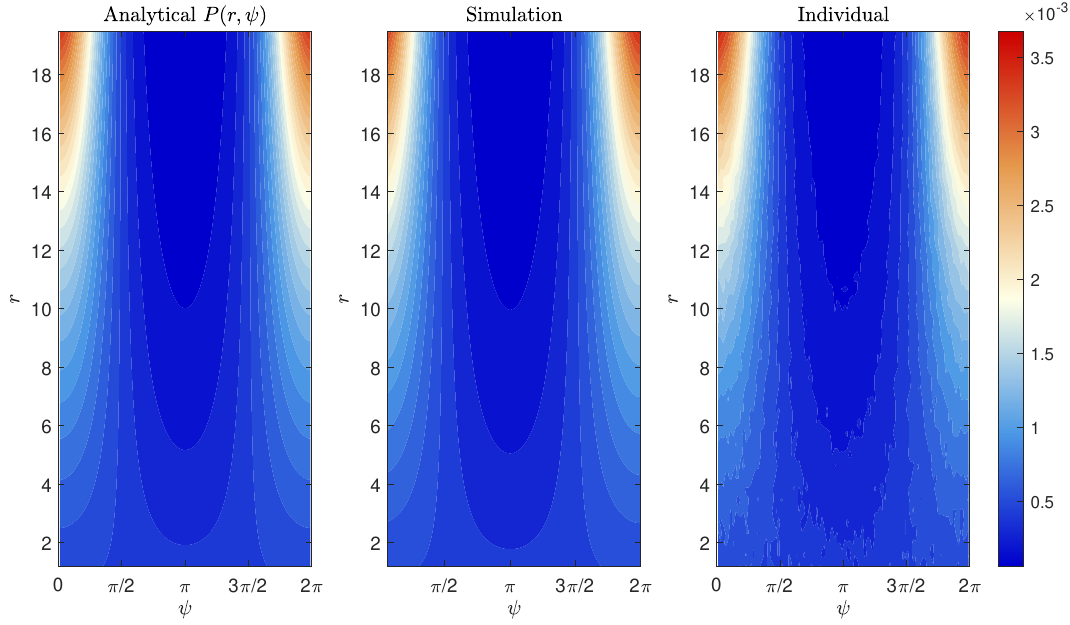}
	\caption{
		\textbf{Contours of exact solution of the Smoluchowski equation without rotational diffusion or background flows validated by stochastic dynamics.}
		In the individual-based model, $10^7$ agent particles are released into a two-dimensional circular domain. The simulation uses a dimensionless time step of $\mathrm{d}t=10^{-2}$ and runs for a total duration of $2000$ time units, sufficient to ensure steady state.
		A fine spatial grid is employed to compute probability distributions and their contour maps.
		The analytical solution shows excellent agreement with stochastic dynamics results.
	}
	\label{EDFig4}
\end{figure*}

\newpage
\begin{figure*}[ht!]
	\centering
	\includegraphics[width=0.75\linewidth]{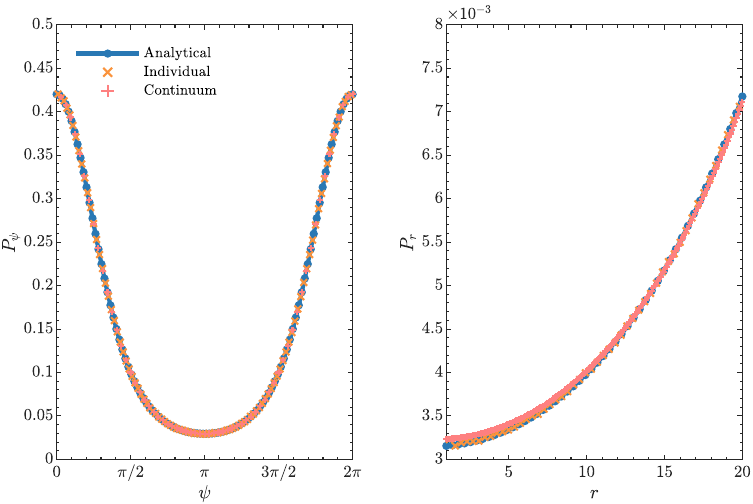}
	\caption{
		\textbf{Orientational and radial distributions without rotational diffusion or background flows from exact solution $P_a$, stochastic dynamics and finite difference simulation.}
		The analytical solution shows excellent agreement with numerical results.
		The numerical details are as in \textbf{Extended Data} \cref{EDFig4}.
	}
	\label{EDFig5}
\end{figure*}

\newpage
\begin{figure*}[ht!]
	\centering
	\includegraphics[width=0.9\linewidth]{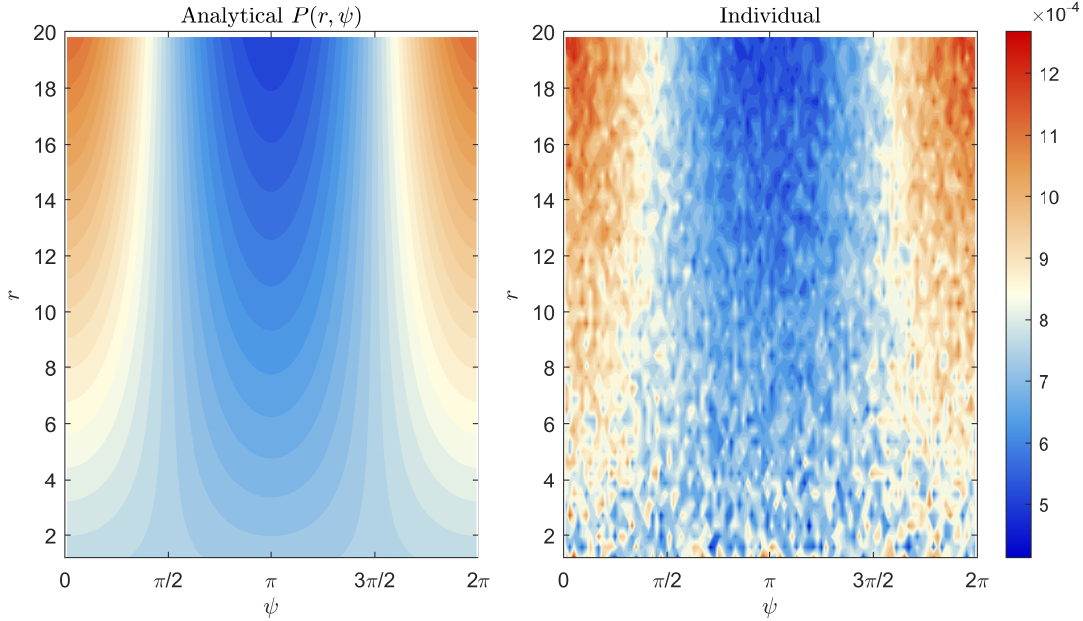}
	\caption{
		\textbf{Exact solution of the Smoluchowski equation without rotational diffusion or background flows validated by stochastic dynamics.}
		In the individual-based model, $10^6$ agent particles are released into a two-dimensional circular domain. The simulation uses a dimensionless time step of $\mathrm{d}t=10^{-1}$ and runs for a total duration of $2000$ time units, sufficient to ensure steady state.
		The analytical solution shows nice agreement with stochastic dynamics results.
		A finer spatial grid can be employed for better visualization and comparison with regard to contour maps, see e.g. setup in \textbf{Extended Data} \cref{EDFig4}.
	}
	\label{EDFig6}
\end{figure*}

\renewcommand{\theequation}{S\arabic{equation}}
\renewcommand{\thetable}{S\arabic{table}}
\setcounter{equation}{0}
\renewcommand{\thefigure}{S\arabic{figure}}
\setcounter{figure}{0}
\renewcommand{\thesection}{\arabic{section}}
\setcounter{section}{0}

\renewcommand{\figurename}{Supplementary Fig.}
\renewcommand{\tablename}{Supplementary Table}

\newpage

\onecolumngrid

\setcounter{page}{1}
\renewcommand{\thepage}{S\arabic{page}}
\renewcommand{\citenumfont}[1]{S#1}
\renewcommand{\bibnumfmt}[1]{[S#1]}

\begin{center}
	
	{\large \bfseries
		Supplementary Information for\\[0.5em]
		Active phase-space topology unifies depletion and alignment in bacterial flows
		\par}
	
	\vspace{1.5em}
	
	Mingyang Guan$^{1}$,
	Bowen Ling$^{1,2}$,
	Enhao Liu$^{3,4,5}$,
	Guoqian Chen$^{6,7}$,
	and Zhan Wang$^{1,2,*}$
	
	\vspace{1em}
	
	{\small
		$^{1}$Key Laboratory for Mechanics in Fluid Solid Coupling Systems, Institute of Mechanics, Chinese Academy of Sciences, Beijing 100190, China\\
		$^{2}$School of Engineering Science, University of Chinese Academy of Sciences, Beijing 100049, China\\
		$^{3}$State Key Laboratory of Deep Earth Processes and Resources, Guangzhou Institute of Geochemistry/Guangdong Provincial Key Laboratory of Mineral Physics and Materials, Guangzhou Institute of Geochemistry, Chinese Academy of Sciences, Guangzhou 510640, China\\
		$^{4}$Guangdong Research Centre for Strategic Metals and Green Utilization, Guangzhou 510640, China\\
		$^{5}$University of Chinese Academy of Sciences, Beijing 100049, China\\
		$^{6}$Department of Mechanics, Peking University, Beijing 100871, China\\
		$^{7}$Key Laboratory of River Basin Digital Twinning of Ministry of Water Resources, National Observation and Research Station of Coastal Ecological Environments in Macao, Macau University of Science and Technology, Macao SAR 999078, China
	}
	
	\vspace{1em}

\end{center}

\vspace{2em}

\begin{itemize}
	\item Supplementary Note 1. Unified Smoluchowski model and analytical solutions\dotfill S2
	\item Supplementary Note 2. 3D vortical flows near a flat plate and local azimuthal drift\dotfill S4
	\item Supplementary Note 3. Generalized von Mises solutions for passive scalars and no-flow cases\dotfill S5
	\item Supplementary Note 4. Dynamical origin of depletion and rheotaxis\dotfill S7
	\item Supplementary Note 5. Conserved motion and organized phase-space separatrices\dotfill S8
	\item Supplementary Table 1. Parameters used in simulations and experiments\dotfill S11
	\item Supplementary Videos 1--5\dotfill S12
	\item Supplementary Figure S1. Comparison of no-flow benchmark with simulations and continuum theory\dotfill S13
	\item Supplementary Figure S2. Shear compression of separatrices in various shear strengths\dotfill S14
	\item Supplementary Figure S3. Experimental observation of trapped \textit{E. coli}\dotfill S15
	\item Supplementary Figure S4. Reproducible vortex-driven trapping of \textit{S. pasteurii}\dotfill S16
	\item Supplementary References\dotfill S17
\end{itemize}

\vspace{150pt}
\noindent\rule{0.1\linewidth}{0.4pt}

\noindent{\large $^*$ \small \textcolor{magenta}{zwang@imech.ac.cn}}

\newpage
\section*{Supplementary Note 1. Unified Smoluchowski model and analytical solutions}\label{SN1}

We consider dilute suspensions of self-propelled elongated swimmers in two-dimensional incompressible flows. 
Each swimmer is characterized by its position $\bm{r}$ and orientation $\bm{p}$, whose stochastic dynamics are governed by
\begin{equation}
	\dot{\bm{r}}=\bm{v}_f(\bm{r})+V_s\bm{p} +\bm{\xi}_r,
	\qquad
	\dot{\bm{p}}
	=\bm{\Omega}(\bm{r},\bm{p}) \bm{\times} \bm{p} + \bm{\xi}_p,
\end{equation}
where $\bm{v}_f$ is the imposed flow, $V_s$ is the swimming speed, and $\bm{\xi}_r$ and $\bm{\xi}_p$ are Gaussian white noises.
The angular velocity takes the Jeffery form
$	\bm{\Omega}(\bm{r},\bm{p}) = \bm{\Omega}_f(\bm{r},\bm{p})/2 + B \bm{p} \bm{\times} \left[\bm{E} {\cdot} \bm{p} \right],$
where $\bm{\Omega}_f=\bm{\nabla} \bm{\times} \bm{v}_f$ is the local flow vorticity, $\bm{E}= (\bm{\nabla}\bm{v}_f+\bm{\nabla}\bm{v}_f^{\text{T}})/2$ is the rate-of-strain tensor, and $B= (\Lambda^2-1)/(\Lambda^2+1)$ is the Bretherton parameter that describes particle shape, with $\Lambda$ the aspect ratio.

The corresponding Smoluchowski equation for the steady-state probability density $P(\bm{r},\bm{p})$ reads
\begin{equation}  
	\bm{\nabla}_{\bm{r}} {\cdot} \bm{J}_r
	+
	\bm{\nabla}_{\bm{p}} {\cdot} \bm{J}_p
	=
	0,
	\label{eq:Smol}
\end{equation}
where $\bm{J}_r=(\bm{v}_f+V_s\bm{p}) P - D_t \bm{\nabla}_r P$ and $\bm{J}_p= \dot{\bm{p}} P - D_r \bm{\nabla}_p P$ are respective flux density tensors in the position and orientation space, $D_t$ and $D_r$ are respective translational and rotational diffusivities,
supplemented by normalization
\begin{eqnarray}
	\int_{\bm{r}} \int_{\bm{p}} P(\bm{r},\bm{p})~\mathrm{d} \bm{r}~\mathrm{d} \bm{p} =1.
\end{eqnarray}

We decompose the phase space into infinite global spaces $\bm{Q}$ and confined local spaces $\bm{q}$, such that the Smoluchowski equation is rewritten as
\begin{eqnarray}
	\mathcal{L}_{Q} P
	+
	\mathcal{L}_{q} P
	=
	0,
\end{eqnarray}
where $\mathcal{L}_Q = \bm{\nabla}_{\bm{Q}} {\cdot} \bm{J}_Q$ and $\mathcal{L}_q=\bm{\nabla}_{q} {\cdot} \bm{J}_q$ are the equivalent operators, with flux density tensors $\bm{J}_Q$ and $\bm{J}_q$ in the global and local spaces, respectively.
The global space and corresponding variables can be multi-dimensional.

Define local moments
\begin{eqnarray}
	P_k (\bm{q}) = \int_{-\infty}^{\infty} Q^k P(Q,\bm{q}) ~\mathrm{d} Q,\quad k=0,1,2,\ldots,
\end{eqnarray}
where $Q$ is the corresponding variable (single in the present work) in the global space. 
We derive the hierarchy moment equations as
\begin{eqnarray}
	\mathcal{L}_{q} P_k = k U_Q P_{k-1} + k(k-1) D_t P_{k-2}, \quad k=0,~1,~2,\ldots,
\end{eqnarray}
where $U_Q$ is the total velocity along $Q$ and $P_{-2}=P_{-1}=0$.
In general, the $n$-th order local moment can be expanded as \citep{bartonMethodMomentsSolute1983,jiangTransientDispersionProcess2021}
\begin{eqnarray}
	P_k(\bm{q},t) =  \sum\limits_{i=1}^\infty p_{ki}(t) \exp(-\lambda_i t) g_i(\bm{q}), 
\end{eqnarray}
where $\lambda_i$ and $g_i$ are eigenvalues and eigenfunctions for the local operator $\mathcal{L}_q$, and $p_{ni}$ is the coefficients to be determined by initial conditions.
In the main text only the steady-state zeroth moment $P_0$ is required; higher moments follow successively.
For asymptotically long times, $P_0$ asymptotes to $P$ through homogenization \citep{guanMigrationConfinedMicroswimmers2024}.
For this eigenvalue problem, $P_0$ can be obtained through eigenfunction expansions with bi-orthogonal basis function $\left\{g_j\right\}_{j=1}^{\infty}$ and a coefficient vector $\bm{c}$ determined by the bilinear equation corresponding to $\mathcal{L}_q$ and the normalization condition.

The Smoluchowski equation for confined active particles takes the form of a drift–diffusion equation with activity-dependent boundary conditions. 
In contrast to classical solute transport problems, the local operator is non-self-adjoint due to active swimming and Robin-type confinement, so standard separation of variables and Sturm–Liouville theory cannot be directly applied \citep{jiangDispersionActiveParticles2019,jiangTransientDispersionProcess2021,wangCrosschannelDistributionStreamwise2022,caraglioAnalyticSolutionActive2022,guanPreasymptoticDispersionActive2023}. 
For Robin boundary conditions, we have
\begin{eqnarray}
	\bm{J}_r {\cdot} \bm{n} =0,
\end{eqnarray}
where $\bm{n}$ is the unit outward normal vector on the boundary $z=\pm W/2$.
The eigenfunctions for Robin boundary conditions in confined Poiseuille flows are solved analytically as
\begin{eqnarray}
	\dfrac{P_R}{\sqrt{2\pi}},~
	\dfrac{P_R}{\sqrt{\pi}} \cos(\dfrac{n\pi z}{W}),~
	\dfrac{P_R}{\sqrt{\pi}} \cos(m \psi),~
	\dfrac{P_R}{\sqrt{\pi}}\sin(m \psi),~
	P_R \sqrt{\dfrac{2}{\pi}}\cos(\dfrac{n\pi z}{W})\cos(m \psi),~
	P_R \sqrt{\dfrac{2}{\pi}}\cos(\dfrac{n\pi z}{W})\sin(m \psi),
\end{eqnarray}
where $P_R=\exp \big[ V_s \sin\psi (z-W/2) /D_t\big]$ is an exact solution for zero rotational diffusion and flow strength.

In confined two-dimensional rotlet flows $r\in [1,r_m]$, the eigenfunctions for Robin boundary conditions are solved analytically as
\begin{eqnarray}
	\dfrac{P_a}{\sqrt{2\pi}} R_0(r),~
	\dfrac{P_a}{\sqrt{\pi}} \cos(m \psi) R_0(r),~
	\dfrac{P_a}{\sqrt{\pi}} \sin(m \psi) R_0(r),~
	\dfrac{P_a}{\sqrt{2\pi}} R_n(r),~
	\dfrac{P_a}{\sqrt{\pi}} \cos(m \psi) R_n(r),~
	\dfrac{P_a}{\sqrt{\pi}} \sin(m \psi) R_n(r),
\end{eqnarray}
where $P_a=\exp \big( V_s \cos\psi~r /D_t\big)$ is an exact solution for zero rotational diffusion and flow strength (see \textbf{Extended Data Figs.}~4 to 6 for numerical validation), and the radial eigenfunctions are 
\begin{eqnarray}\label{eq_eigenfunction_radial}
	R_0(r) = \sqrt{\dfrac{2}{r_m^2-1}},\quad
	R_n(r) = D_n \left[ Y_1(k_n) J_0(k_n r) - {J_1(k_n)} Y_0(k_n r) \right],\quad n=1,2,3,\ldots,
\end{eqnarray}
where $J_0$ and $Y_0$ are zeroth order Bessel functions of the first and second kind, and $J_1$ and $Y_1$ are first order Bessel functions of the first and second kind.
The wavenumbers $k_n$  are the positive roots of the transcendental equation
\begin{equation}
	\dfrac{\mathrm{d}}{\mathrm{d}r} R_n(r_m)=0 \Rightarrow
	J_1(k_n) Y_1(k_n r_m) - J_1(k_n r_m) Y_1(k_n) = 0, \quad n = 1, 2, \ldots
\end{equation}
which ensure the Neumann boundary conditions
\begin{eqnarray}
	\left. \dfrac{\mathrm{d} R}{d r}\right|_{r=1}=0,\quad
	\left. \dfrac{\mathrm{d} R}{d r}\right|_{r=r_m}=0.
\end{eqnarray}
The normalization constant $D_n$ is chosen so that:
\begin{equation}
	\int_1^{r_m} R_n^2(r)~r~\mathrm{d}r = 1.
\end{equation}
The steady state is constructed by factoring out the exact advective equilibrium, reducing the problem to a Neumann Laplacian eigen-expansion for the residual field. 
The resulting Robin basis naturally encodes the coupling between position and orientation.

For periodic boundary conditions where anti-symmetric parabolic flow profiles repeat periodically with alternating sign \citep{rusconiBacterialTransportSuppressed2014}, we impose natural conditions in $\psi$ as
\begin{eqnarray}
	\left. P\right|_{\psi=0} = \left. P\right|_{\psi=2\pi},\quad
	\left. \frac{\partial P}{\partial \psi}\right|_{\psi=0} = \left. \frac{\partial P}{\partial \psi}\right|_{\psi=2\pi}
\end{eqnarray}
and periodicity at the walls $z=\pm W/2$.
The eigenfunctions for these anti-symmetric boundary conditions are solved analytically as
\begin{eqnarray}
	&\dfrac{1}{\sqrt{2\pi}},\quad \dfrac{1}{\sqrt{\pi}} \sin (m\psi),\quad
	\dfrac{1}{\sqrt{\pi}} \cos (m\psi),\quad \dfrac{1}{\sqrt{\pi}} \cos(\dfrac{2n\pi z}{W}),\quad \dfrac{1}{\sqrt{\pi}} \sin(\dfrac{2n\pi z}{W}),\nonumber\\
	&\sqrt{\dfrac{\pi}{2}} \cos(\dfrac{2n\pi z}{W}) \sin(m\psi),~
	\sqrt{\dfrac{\pi}{2}} \cos(\dfrac{2n\pi z}{W}) \cos(m\psi),~
	\sqrt{\dfrac{\pi}{2}} \sin(\dfrac{2n\pi z}{W}) \sin(m\psi),~
	\sqrt{\dfrac{\pi}{2}} \sin(\dfrac{2n\pi z}{W}) \cos(m\psi),
\end{eqnarray}
with $n=1,~2,~\ldots$ and $m=1,~2,~\ldots$.

\clearpage

\newpage
\section*{Supplementary Note 2. 3D vortical flows near a flat plate and local azimuthal drift}\label{SN4}

\textbf{3D vortical flows generated by a rotating sphere near a flat plate.}
We consider a rigid sphere of radius $R$ rotating with a constant angular velocity $\bm{\omega}$ in an otherwise quiescent, incompressible Newtonian fluid of viscosity $\mu$.
At low Reynolds numbers,
$
\mathrm{Re}={\rho\,\omega R^2}/{\mu}\ll 1,
$
the flow obeys the steady Stokes equations
$
-\nabla p+\mu\nabla^2\bm{u}=0,~
\nabla\cdot\bm{u}=0,
$
subject to no–slip at the particle surface and vanishing velocity at infinity,
$
\bm{u}(r=R)=\bm{\omega}\times\bm{r},~
\bm{u}(r\to\infty)=0.
$
Because the sphere undergoes pure rotation without translation, the induced flow is purely toroidal. By symmetry, the velocity field can be written in the rotlet form
$
\bm{u}(\bm{r})=f(r)\,(\bm{\omega}\times\bm{r}),
$
which is divergence-free by construction. The anti-symmetry of the flow implies a spatially uniform pressure, which may be set to zero. Substitution into the Stokes equation yields
$
\nabla^2\bm{u}=0,
$
leading to the ordinary differential equation
$
f''(r)+{4}f'(r)/{r}=0.
$
Imposing decay at infinity and no-slip at $r=R$ gives
$
f(r)={R^3}/{r^3},
$
so that the velocity field reads
\begin{eqnarray}
	\bm{u}(\bm{r})
	=\frac{R^3}{r^3}\,(\bm{\omega}\times\bm{r}).
\end{eqnarray}
This is the classical rotlet flow, corresponding to the fundamental Stokes solution generated by a point torque. 
The flow is purely azimuthal, decays as $u\sim r^{-2}$, and possesses non-zero vorticity,
$
\nabla\times\bm{u}
=
2{R^3}\bm{\omega}/{r^3}.
$

\textbf{Quasi-two-dimensional reduction and image corrections.}
In the experiments, particle motion is effectively confined to a thin horizontal layer. Restricting attention to the mid-plane and assuming the rotation axis perpendicular to the plane, the flow reduces to a quasi-2D azimuthal field,
$
\bm{u}
=
u_\phi(r)\,\hat{\bm{e}}_\phi,~
u_\phi(r)\simeq {R^3\omega}/{r^2}.
$
The reduced decay $u_\phi\sim r^{-2}$ reflects the loss of one spatial dimension upon vertical averaging.

The presence of a solid bottom wall enforces a no-slip condition. To leading order, this effect can be incorporated through a hydrodynamic image system.
In the far field ($r\gg R$), the dominant correction arises from a mirror rotlet of opposite sign located at a distance $2R$ below the wall, yielding the approximate form
\begin{eqnarray}
	v_f(r)
	\simeq
	\frac{R^3\omega}{r^2}
	-
	\frac{R^3\omega\, r}{\bigl[r^2+(2R)^2\bigr]^{3/2}},
\end{eqnarray}
where the second term cancels the azimuthal velocity at the wall to first order. 
Higher-order singularities (e.g., stresslets) become relevant only at distances comparable to the particle radius.

Introducing the intrinsic Stokes scales for length and velocity,
$
\tilde r={r}/{R},~
\tilde u={u}/{R\omega},
$
the wall-corrected Stokes solution collapses onto a parameter-free form as ${1}/{\tilde r^2}-{\tilde r}/{(\tilde r^2+4)^{3/2}}$.
As a consequence, azimuthal velocity profiles measured for different vortex-core radii collapse onto a single master curve when rescaled by $R$ and $R\omega$, in excellent agreement with experiments (see \textbf{Fig.}~2 in the main text).

\textbf{Mean azimuthal drift with orientational locking to the background flow.}
A central experimentally accessible quantity is the mean azimuthal velocity of bacterial trajectories around the vortex centre. 
From particle tracking, this corresponds to the lab-frame azimuthal drift $\dot\phi$. The local drift velocity of swimmers can be written as
$
\dot\phi= \Omega_\phi(r)+{V_s}\sin\psi/{r},
$
where $\Omega_\phi(r)= v_f / r$ is the azimuthal velocity of fluid particles.
The mean azimuthal drift at a given radius $r$ is therefore
\begin{eqnarray}
	\langle\dot\phi\rangle(r)
	=
	\int_0^{2\pi}
	\left[
	\Omega_\phi(r)
	+
	\frac{V_s}{r}\sin\psi
	\right]
	P(r,\psi)\,\mathrm d\psi,
\end{eqnarray}
where $P(r,\psi)$ is the steady-state joint distribution of radial position and relative orientation.

Remarkably, in the limit of strongly elongated (rod-like) axis-symmetric swimmers ($B\to1$), the angular drift is exactly locked to the background vortical flow,
$
\langle\dot\phi\rangle(r) = \Omega_\phi(r),
$
independent of swimming speed $V_s$ and noise strengths, provided the system reaches a steady state.
This result follows from a vanishing angular probability flux in the steady-state relative-orientation dynamics.

As shown in \textbf{Fig.} 4a, cells tend to align near $\psi=\pi/2$ and $\psi=3\pi/2$, albeit with a small systematic deviation from these values. This behaviour can be rationalized by analyzing the strong-rotation limit of the governing Smoluchowski dynamics. If the angular speed of rotating sphere goes to infinity ($\omega \to \infty$), the stationary solution of the angular equation tends to
$
P={\sqrt{1-B^2}}/{2 \pi \left(1+B \cos 2\psi\right)},~ B<1,
$
which generalizes the von Mises distribution and is peaked at the minima of $1+B \cos 2\psi$. For spherical micro-swimmers ($B=0$), the distribution becomes uniform, whereas in the singular slender limit $B=1$, the dominant balance reduces to
$
{\mathrm{d}}/{\mathrm{d}\psi}\!\left[(1+\cos 2\psi)P\right]=0,
$
whose solution diverges at $\cos \psi=0$. Consequently, the distribution collapses to a pair of delta functions:
$
P(\psi) \rightarrow \left[\delta\!\left(\psi-{\pi}/{2}\right)+\delta\!\left(\psi-{3\pi}/{2}\right)\right]/2.
$
In the vortex-dominated regime, swimmers preferentially align perpendicular to the principal stretching direction, corresponding to stagnation points of the angular drift.
Hence, at the leading order of weak anisotropy and activity, $P(r,\psi)$ is even in $\psi$ and $\sin\psi$ is anti-symmetric such that
$
\int_0^{2\pi}\sin\psi\,P(r,\psi)\,\mathrm d\psi = 0.
$
Therefore, active swimming does not generate any systematic correction to the mean azimuthal advection. 
While activity strongly modifies orientation statistics and angular fluctuations, it cannot renormalize the mean angular drift in this idealised rod-like limit. 
For finite shape anisotropy ($B<1$) or strong noise, small deviations from perfect locking may arise, but the leading-order behaviour remains dominated by the background flow.

The exact drift locking, 
\begin{eqnarray}
	\tilde u(\tilde r) = \langle\dot\phi\rangle~\tilde{r} \simeq \frac{1}{\tilde r^2}
	-\frac{\tilde r}{(\tilde r^2+4)^{3/2}}.,
\end{eqnarray}
is closely analogous to the classical Taylor–Aris dispersion in shear flows, where transverse diffusion and shear dramatically enhance streamwise dispersion while leaving the mean drift fixed by the imposed flow. 
Here, swimming-induced orientational dynamics plays the role of an additional transverse stochastic process: it enhances angular dispersion and residence-time fluctuations, but the mean azimuthal motion of bacterial trajectories remains determined solely by the vortex flow to the first order.

\section*{Supplementary Note 3. Generalized von
	Mises solutions: passive scalars in rotlet flow and micro-swimmers without flow}\label{SN3}

For passive particles $V_s=0$ in a two-dimensional rotlet flow, the steady-state Smoluchowski equation can be reduced to
$
\partial_\psi
\left[{\omega}(1+B\cos2\psi)P/{r^2} \right]+\left(1+{D_t}/{r^2}\right) {\partial_{\psi\!\psi}}P=0.
$
This equation still has $r$-dependence through $\omega/r^2$, which acts as a constant at fixed $r$.
Let the probability flux be constant or the angular integration with regard to $\psi$ gives
$	
\left(1+{D_t}/{r^2}\right) \partial_\psi P= - {\omega}	(1+B\cos2\psi)P/{r^2},
$
with general solutions of the form
\begin{eqnarray}
	\ln P = -\dfrac{\omega}{r^2+D_t} \left(\psi + \frac{B}{2} \sin 2\psi\right).
\end{eqnarray}
To ensure periodicity in $\psi$, we give an approximate solution as
\begin{eqnarray}
	P(r,~\psi) \sim \exp{\left(-\dfrac{\omega B}{2(r^2+D_t)}\sin2\psi\right)},\label{benchmark_no_activity}
\end{eqnarray}
or, with proper normalization
\begin{eqnarray}
	\int_1^{r_m}\int_0^{2\pi} P~ r~\mathrm{d}\psi\mathrm{d}r=1. \label{eq_SN_normalization}
\end{eqnarray}
This benchmark solution of $P$ is periodic and of physical significance.

This steady-state distribution has the form of a skewed von Mises–like function, shaped by the cosine term arising from Jeffery torques in micro-swimmer dynamics. 
The distribution peaks at orientations where $\sin 2\psi = -1$, namely $\psi= 3\pi/4$ and $\psi=7\pi/4$.
These correspond to directions diagonally misaligned with the flow ---  at $45^\circ$ counterclockwise from the flow and clockwise from the backward flow.
Physically, the shape-induced torque $1 + B \cos 2\psi$ rotates particles at different angular velocities depending on their orientation.
The steady-state distribution emerges from a balance of
(a) slow rotation near orientations that resist torque (where particles linger longer),
(b) rapid rotation near unstable orientations (particles quickly rotate through),
and (c) rotational diffusion which smooths the distribution.
As a result, particles do not align exactly with the flow. Instead, their orientations are biased toward diagonally offset angles, due to the angular dependence of the torque.

As the prefactor $\omega B / r^2$ increases, the orientation distribution becomes increasingly sharp.
In the limit $\omega B \gg 1$, Eq.\ (2)
becomes tightly concentrated near $\psi = 3\pi/4$ and $7\pi/4$ corresponding to alignment with the flow direction. However, the distribution never becomes a perfect delta function unless rotational diffusion is entirely absent.
On the other hand, the solution fails to capture the main balance mechanism for large $\omega$ because the angular profile becomes sharply peaked due to the steep effective potential $U \sim \omega B \sin 2\psi/ \big[2(r^2+D_t)\big]$.
Moreover, as $\omega$ increases, any mismatch between the drift and diffusion operators becomes amplified; the radial diffusion term becomes non-negligible again because the solution has steep angular gradients that indirectly affect radial profiles (e.g., via torque-induced coupling).

The dominance of $\omega$ arises from its role in governing the strength of angular advection relative to angular diffusion. In the Smoluchowski equation, the angular drift term scales as $\omega(1 + B \cos 2\psi)/r^2$, while angular diffusion scales as $1 + D_t/r^2$. Their ratio defines an effective angular Péclet number, $ \mathrm{Pe}_\psi = \omega B / (r^2 + D_t)$, which controls the sharpness of the angular distribution. The benchmark solution 
$
P(r,\psi) \propto \exp(-{\mathrm{Pe}_\psi} \sin 2\psi / {2}),
$
is derived by assuming a quasi-equilibrium in $\psi$, balancing drift and diffusion.

As $\omega$ increases, $\mathrm{Pe}_\psi$ becomes large, leading to sharply peaked angular distributions that cannot be captured by the single-mode exponential ansatz. In contrast, increasing $D_t$ reduces $\mathrm{Pe}_\psi$, flattening the angular distribution, and making the ansatz more accurate. Therefore, the accuracy of the approximate solution is primarily limited by $\omega$, not $D_t$, because it directly sets the nonlinearity in $r$ and anisotropy in $\psi$ of the angular flux that the ansatz must approximate.

For the special case without background flows $\omega=0$, if we assume a small net radial flux for asymptotically long times, the remaining angular governing equations become
\begin{eqnarray}
	\left(1+\dfrac{D_t}{r^2}\right) \dfrac{\partial^2 P}{\partial \psi^2}
	+\dfrac{V_s}{r}\dfrac{\partial \left(\sin\psi P\right)}{\partial \psi} =0.
\end{eqnarray}
A 1D benchmark solution takes the form 
\begin{eqnarray}
	P(\psi) \propto \exp{\left[\dfrac{V_s r}{(D_t+r^2)} \cos\psi \right]},
	\label{eq_1d_sol_no_flow}
\end{eqnarray}
with proper normalization as \cref{eq_SN_normalization}.
Note that this solution violates the Robin boundary condition, and may fail to describe the rotational dynamics when wall accumulation is dominant.
The approximation assumes (i) small radial flux everywhere and (ii) mild $r$-dependence of the angular concentration. 
Both fail once $V_s$ is large compared to the radial diffusion/geometry scale, so the full $r$–$\psi$ coupling and boundary layers matter.
However, it could be a nice approximation to the orientational distribution averaged over $r$ (see \textbf{Supplementary} \cref{figS1}), where radial nonuniformity near the walls is smoothed through averaging.

The assumptions behind \cref{eq_1d_sol_no_flow}
break down at large swimming P\'{e}clet for the following coupled reasons:
(a) The neglected coupling term grows with $\kappa$.
With $P\propto \exp\big[{\kappa(r)\cos\psi}\big]$, we have $
\partial_r \ln P=\kappa'(r)\cos\psi,~ \kappa'(r)={V_s\,(D_t-r^2)}/{(D_t+r^2)^2}.
$
The residual of the full equation when we plug this benchmark back in is
$\mathcal{R}
={V_s\cos\psi}\partial_r(rP)/{r}-{D_t}\partial_r\!\left(r\partial_r P\right)/{r},
$
because the angular part cancels by construction. Using the expressions above,
$
\mathcal{R}\sim P\big\{
{V_s\cos\psi}\big[1+r\kappa'(r)\cos\psi\big]/{r}
-{D_t}\big[\kappa'(r)\cos\psi + r(\kappa''(r)\cos\psi+(\kappa'(r))^2\cos^2\psi)\big]/{r}
\big\}.
$
For $|\kappa|\gg1$ (large $V_s$ and/or small $D_t$ over parts of the domain), the terms $\propto \kappa',(\kappa')^2,\kappa''$ become $O(\kappa)$ or larger, so the discarded radial balance is no longer small.
(b) Wall accumulation creates boundary layers (Robin boudnary conditions violated).
At large $V_s$, orientations concentrate near $\psi=0$ (outward) and $\psi=\pi$ (inward). The $\cos\psi$ radial drift then pushes probability to the walls, and a non-negligible radial diffusive flux is required to balance it. That means sharp radial boundary layers form near the walls, which the ansatz cannot capture (and recall that it indeed violates the Robin condition).
(c) Strong $r$-variation of $\kappa(r)$ spoils a single “angular-only” structure.
$\kappa(r)=V_s r/(D_t+r^2)$ varies across the gap and even changes slope at $r=\sqrt{D_t}$. For large $V_s$, the orientational distribution varies dramatically with $r$, so $r$-averaging no longer “smooths out” wall effects; instead it mixes very different local von Mises shapes, producing visible deviations from a single von Mises profile.

\newpage
\section*{Supplementary Note 4. Dynamical origin of depletion and rheotaxis}\label{SN2}
A quantitative comparison between the Langevin theory and experiments in planar Poiseuille flows \citep{rusconiBacterialTransportSuppressed2014} demonstrates that wild-type \textit{B.~subtilis} develop robust centreline depletion. 
Our minimal model captures both the magnitude and shear-rate dependence of this depletion across three bacterial species (namely wild-type \textit{B.~subtilis}, smooth-swimming \textit{B.~subtilis} and wild-type \textit{P.~aeruginosa})(\textbf{Fig. 3a}).
Using a single parameter set ($V_s,~B$) calibrated to wild-type B. subtilis, the model reproduces main features of the measured probability distributions, with residual discrepancies naturally explained by species-dependent swimming speeds and Bretherton parameters not available in the literature. 
This agreement indicates that depletion is a generic outcome of active transport in shear, rather than a species-specific biological strategy.

As the shear increases in Poiseuille flows, centreline depletion weakens and smoothly evolves into a symmetric double-peaked profile (\textbf{Fig. 3a}, right). 
This evolution does not reflect a change in transport physics, but instead arises from the progressive compression of separatrices in the phase-space (\textbf{Supplementary Fig.} S2), which shortens swimmer residence times near the centreline and redistributes toward off-centre regions.

The underlying dynamics are further exposed by the steady-state joint distribution of position and orientation $P(z,\psi)$ (\textbf{Fig. 1b}), where the orientation angle $\psi$ is measured relative to the flow (\textbf{Fig. 3b}).
The individual-based model, equation (3), with antisymmetric periodic boundaries reproduces the experimental statistics \citep{rusconiBacterialTransportSuppressed2014} satisfactorily, indicating that bulk statistics are not sensitive to boundary details; imposing no-flux Robin conditions quantitatively alters the depletion location. 
Notably, depletion is invariably accompanied by strong rheotactic alignment: upstream-oriented states dominate, while downstream orientations are slightly suppressed. 
This suggests that depletion and rheotaxis in fluid shear are not independent phenomena but arise simultaneously from the same phase-space structures.

Experiments on bacteria in 2D rotlet flows reveal two striking features: a strong orientational bias and a rapid expulsion of swimmers from the vortex core \citep{sokolovRapidExpulsionMicroswimmers2016}. 
The laboratory frame is defined in polar coordinates $(r,~\phi)$ and the particle orientation is characterized by a relative angle measured with regard to the local radial direction, see \textbf{Fig. 1c}.
As shown in \textbf{Fig. 4a}, bacteria preferentially align near $\psi=\pi/2$ and $\psi=3\pi/2$, corresponding to swimming with or against the local azimuthal flow.
This alignment is systematically offset from exact streamline-following, indicating an intrinsic rheotactic bias rather than passive advection.

This behaviour follows naturally from the strong-vortex limit of the Smoluchowski description. 
As the vortex strength $\omega$ becomes large, the steady-state orientation distribution converges to a generalized von Mises form
$
P={\sqrt{1-B^2}}/ \big[{2 \pi \left(1+B \cos 2\psi\right)}\big],~ B<1,
$
with extrema are set by the minima of an effective, shape-induced orientational drift $(1+B\cos2\psi)$ (see \textbf{Supplementary Note} 3). 
Physically, this drift originates from the coupling between shape anisotropy and shear.
Spherical swimmers ($B=0$) lack this coupling and therefore remain isotropically oriented, consistent with the absence of rheotaxis in isotropic particles.
In contrast, rod-like swimmers ($B\to1$) experience a strong bias where rotational flux is suppressed near $\cos\psi=0$, causing orientations to collapse onto two preferred directions. 
These correspond to stagnation points of the deterministic orientation dynamics, where shear-induced rotation vanishes and swimmers become transiently locked swimming either upstream or downstream.

This orientational locking directly impacts bacterial transport. 
Crucially, it does not rely solely on local shear, but instead emerges from the phase-space structure of the coupled position–orientation dynamics. 
In vortical flows, biased orientation suppresses cross-stream motion, leading to efficient depletion from the vortex core.
From a biological perspective, such a mechanism may facilitate escape from recirculating regions in turbulent or vortical environments, promoting dispersal toward new, potentially nutrient-rich habitats.
More broadly, it suggests that bacterial shape alone can encode a passive navigational strategy, without requiring active sensing or behavioural control.

We test this mechanism quantitatively using the individual-based model, which incorporates random swimming and hydrodynamic reorientation. 
As shown in \textbf{Fig. 4a} (right), the reproduced orientation distributions closely match experiments across all radii.
To further quantify the agreement, we exploit the exact steady-state solution of equation (5) in the absence of vortex flow and rotational diffusion,
\begin{eqnarray}
	P_a=\exp\left(\frac{{V_s}r \cos\psi}{D_t}\right),
	\label{eq:sol}
\end{eqnarray}
which serves as a benchmark distribution for the no-flux Robin boundary conditions. 
Normalizing the experimental data by $P_a$ collapses the orientation distributions at different radii onto Gaussian profiles. 
Both the Smoluchowski theory and individual-based model reproduce this collapse.
When further rescaled by their peak values, all profiles merge onto a single, radius-independent master curve (\textbf{Fig. 4b}). 
This universal Gaussian highlights a simple organizing principle: in vortex-dominated flows, bacterial orientation statistics obey a modified diffusive scaling that is largely independent of position (\textbf{Fig. 4c}).

\newpage
\section*{Supplementary Note 5. Conserved motion and organized phase-space separatrices}\label{SN2}

We consider the deterministic limit of active particle dynamics in two-dimensional Poiseuille or swirling flows, obtained by setting translational and rotational noise to zero, $D_t = D_r = 0$.
In this limit, trapping and escaping are not stochastic processes but are entirely determined by the phase–sparce structure of the underlying dynamical system. Below we show that both planar Poiseuille trapping and two-dimensional rotlet trapping belong to the same class of conserved systems, allowing for a unified quantitative criterion distinguishing captured from escaping trajectories.

Let $q$ denote a generalized transverse coordinate and $p$ an orientational angle. In both geometries considered here,
\begin{eqnarray}
	(q,p)=(z,\psi)\quad\text{(Poiseuille)},\qquad
	(q,p)=(r,\psi)\quad\text{(rotlet)},
	\label{eq_conserved_variable}
\end{eqnarray}
the deterministic swimmer dynamics may be written in the form
\begin{eqnarray}
	\dot q = \partial_p H(q,p),\qquad
	\dot p = -\partial_q H(q,p),
\end{eqnarray}
where $H(q,p)$ is a conserved quantity along trajectories:
\begin{eqnarray}
	\frac{dH}{dt}=0.
\end{eqnarray}
The explicit expression for $H$ depends on the flow geometry and shape anisotropy $B$, but in both cases it can be constructed analytically. \cref{eq_conserved_variable} implies that the dynamics is integrable and that trajectories are confined to level sets of $H$ in the $(q,p)$ phase space.

The conserved quantity $H(q,p)$ possesses a saddle point $(q_s,p_s)$, corresponding to an unstable fixed point of the deterministic dynamics. 
The level set
\begin{eqnarray}
	H(q,p) = H_c \equiv H(q_s,p_s)
\end{eqnarray}
defines a separatrix in the phase space.
This separatrix divides phase space into two topologically distinct regions:
\begin{enumerate}
	\item Closed orbits ($H < H_c$): Trajectories are bounded in $q$ and correspond to deterministic trapping.
	\item Open orbits ($H > H_c$):	Trajectories exhibit unbounded drift and correspond to escaping.	
\end{enumerate}
This separatrix structure is present both in planar Poiseuille and two-dimensional rotlet flows, despite the different functional forms of $H$.

\textbf{Planar Poiseuille flows.}
For a planar Poiseuille flow $u=U(1-4{z^2}/{W^2})$ with shear $S= \partial_z u=-8zU/W^2$, the steady-state Smoluchowski equation for the distribution $P(\psi, z)$ reads
\begin{eqnarray}
	\dfrac{\partial P}{\partial t}+\big(u-V_s \cos\psi\big)\dfrac{\partial P}{\partial y}
	-V_s \sin\psi \dfrac{\partial P}{\partial z}
	+D_r \dfrac{\partial^2 P}{\partial\psi^2} +\dfrac{S}{2}\dfrac{\partial}{\partial\psi}
	\big[(
	-1+B \cos 2 \psi)P\big]=0,
\end{eqnarray}
with anti-symmetric periodic boundary conditions at $z=\pm W/2$ so that the parabolic flow profile repeats periodically with alternating sign.
For the deterministic limit,  we can follow Zottl \& Stark \citep{zottlNonlinearDynamicsMicroswimmer2012} to derive the constants of motion and separatrix. The coupled deterministic ODEs are
$
\dot z = -V_s\sin\psi,~
\dot\psi = 4{U}\,z\,(1-B \cos2\psi)/{W^2}.
$
Eliminate time
\begin{eqnarray}
	\frac{dz}{d\psi}=\frac{\dot z}{\dot\psi}
	= -\frac{V_sW^2}{4U}\;\frac{\sin\psi}{z(1-B\cos2\psi)}.
\end{eqnarray}
Integrating both sides and noting
$
1-B\cos2\psi = 1-B(2\cos^2\psi-1)=(1+B)-2B u^2,
$
we obtain (for the usual case $B>0$) the anti-derivative in terms of $\operatorname{arctanh}$:
\begin{eqnarray}
	\frac{z^2}{2}
	= -\;K\;\Big[\,\operatorname{arctanh}\!\big(s\cos\psi\big) + C'\Big],
\end{eqnarray}
where
$
s = \sqrt{{2B}/{(1+B)}},~
K = {V_s W^2}/{4U\sqrt{2B(1+B)}}.
$
Equivalently, we write the conserved constant as \citep{zottlPeriodicQuasiperiodicMotion2013a}
\begin{eqnarray}
	h_1(z,\psi)\;=\;\frac{z^2}{2} + K\,\operatorname{arctanh}\!\big(s\cos\psi\big).
\end{eqnarray}
The separatrix in the $(z,\psi)$ plane is the trajectory through the saddle equilibrium at $(z,\psi)=(0,0)$ (or $(0,\pi)$ for the symmetric branch). Evaluate $H$ at the saddle $(0,0)$:
$
H_{\text{saddle}} \;=\; - K\,\operatorname{arctanh}(s).
$
Set $H(z,\psi)=H_{\text{saddle}}$. Rearranging gives the separatrix implicitly:
\begin{eqnarray}
	\frac{z^2}{2}
	= K\Big[\operatorname{arctanh}(s\cos\psi)-\operatorname{arctanh}(s)\Big].
\end{eqnarray}
This formula defines the two branches of the separatrix which emanate from $(0,0)$ (and the symmetric one from $(0,\pi)$ if we replace $\cos\psi$ by $\cos(\psi-\pi)=-\cos\psi$).

Linearize the $(z,\psi)$ system about the equilibrium $(z,\psi)=(0,0)$. For small $\psi$ and small $z$,
$
\dot z \approx -V_s\psi,~
\dot\psi \approx 4{U}\,(1-B)\,z /{W^2}.
$
Linear system matrix ($\dot{a}=A a$) reads
\begin{eqnarray}
	A=\begin{pmatrix}0 & -V_s\\[4pt]
		4\frac{U}{W^2}(1-B) & 0\end{pmatrix}.
\end{eqnarray}
Its eigenvalues satisfy $\lambda^2 = 4{U}V_s (1-B)/{W^2}$, so
$\lambda = \pm \sqrt{4{U}V_s(1-B)/{W^2}}.$

If $1-B>0$ (i.e. the usual case for prolate swimmers with $0\le B<1$), the eigenvalues are real and of opposite sign --- $(0,0)$ (and $(0,\pi)$) are saddle points (hyperbolic). 
That means the rheotactic orientations $\psi=0,\pi$ at the channel centreline are unstable along one direction (they have stable and unstable manifolds: trajectories approach along one direction and depart along the other). 
The separatrix we wrote is exactly the stable/unstable manifold boundary between different qualitative trajectory types (see \textbf{Supplementary} \cref{figS2} for different initial conditions and shear strengths).
If $1-B<0$ then eigenvalues are imaginary, i.e. centre (neutrally oscillatory). 

In summary, for common prolate swimmers ($0<B<1$) the centreline rheotactic orientations $\psi=0$ (upstream) and $\psi=\pi$ (downstream) at $z=0$ are saddle-type equilibria: they are not simply stable attractors. 
A swimmer exactly at $(0,0)$ stays aligned, but any generic small perturbation will typically move it off the centreline along the unstable manifold, so there is no robust rheotactic trapping at these orientations.
The separatrix expression above separates trajectories that loop in $(z,\psi)$ from those that swing/escape; it also fixes where cusps or turning points occur (these come from the separatrix geometry where $\partial z/\partial\psi$ diverges).
If $B\to0$ (spherical swimmer), simplify $s\to0$ and the arctanh expansion recovers classical expressions in Ref. \citep{zottlNonlinearDynamicsMicroswimmer2012}.

\textbf{Rotlet flows.}
In a rotlet flow, using length scale $R$ and time scale $\omega^{-1}$,
and defining the non-dimensional swimming speed
$\alpha={V_s}/{\omega R},$
the dimensionless equations of motion take the form
\begin{eqnarray}
	\begin{aligned}
		\dot r &= \alpha \cos\psi,\\
		\dot\phi &= \dfrac{\alpha}{r}\sin\psi + \dfrac{1}{r^2},\\
		\dot\psi &= -\dfrac{1}{r^2}(1+B\cos2\psi)-\dfrac{\alpha}{r}\sin\psi.
	\end{aligned}
\end{eqnarray}
Specifically for spherical particles ($B=0$), if we define
\begin{eqnarray}
	h = \alpha r \sin \psi + \ln r,
\end{eqnarray}
Conservation requires
\begin{eqnarray}
	\begin{aligned}
		\dot{h}
		&=\alpha \dot{r} \sin\psi+\alpha r \cos\psi \dot{\psi}+ \frac{\dot{r}}{r}
		= \alpha^2\sin\psi\cos\psi+\alpha r\cos\psi (-\frac{1}{r^2}-\frac{\alpha}{r}\sin\psi)+\frac{\alpha \cos\psi}{r}\\
		&= \alpha^2\sin\psi\cos\psi-\frac{\alpha \cos\psi}{r}-\alpha^2\sin\psi\cos\psi+\frac{\alpha \cos\psi}{r}=0.
	\end{aligned}
\end{eqnarray}
Next, we notice that 
\begin{eqnarray}
	\frac{\dot r^2}{2}
	+\frac{\alpha^2}{2}\!\left[
	\left(\frac{h-\ln r}{\alpha r}\right)^2
	-1
	\right]=0.
\end{eqnarray}
Thus $r$ behaves as if it was under the influence of an effective potential
\begin{eqnarray}
	V(r)
	=\frac12\left[
	\left(\frac{h-\ln r}{r}\right)^2
	-\alpha^2
	\right],
\end{eqnarray}
with an energy-like quantity $E= \dot{r}^2/2+V(r) \equiv 0$ for all times. 
Since the effective potential V has limits $V\to -\alpha^2/2$ as $r \to \infty$ and $V\to +\infty$ as $r\to 0$, the entrapment relies on the existence of a local maximum of $V(r)>0$ to prevent the swimmer from escaping to infinity. 
Taking the derivative, we see that the condition $dV/dr=0$ is equivalent to $r=\exp(h)$ or $r=\exp(1+h)$. At $r=\exp(h)$ we have a minimum of the potential, with $V_\text{min}= -\alpha^2/2$ while at $r=\exp(1+h) $ there is a local maximum $V_{\text{max}}=\left(
\textit{e}^{-2(h+1)}-\alpha^2 \right)/2$. Thus, we predict theoretically that the swimmer will be trapped if  and only if $r_0<\exp(-1-h)$ and $V_{\text{max}}>0$ so that $h<-1-\ln\alpha$. The fixed point ($1/\alpha,3\pi/2$) is a saddle point, which determines exactly the maximum radius from which trapping could happen.

For elongated particles ($B>0$), Tanasijevic \& Lauga \citep{tanasijevicMicroswimmersVorticesDynamics2022} obtained a conserved quantity
\begin{eqnarray}
	h_2= \dfrac{\gamma K_0(\eta)+\sin \psi K_1 (\eta)}{\gamma I_0(\eta)-\sin \psi I_1 (\eta)}
\end{eqnarray}
where $\eta=\sqrt{2 B(1+B)}/(V_s r)$, $\gamma=\sqrt{(B+1)/(2B)}$, and $I, K$ are the modified Bessel functions of the first kind. We come to a conclusion that a swimmer will be trapped in a 2D rotlet if and only if initially $r_0<r_c= (1-B)/V_s$ and $h_2(r_0,\psi_0)<h_c(r_c)$, where $h_c={(\gamma K_0- K_1)}/
{(\gamma I_0+ I_1)}$ is the minimum of $h_2$ evaluated at $r=r_c$.

For experimentally relevant parameters as in Ref.~\citep{sokolovRapidExpulsionMicroswimmers2016}, trapping exists only very close to the vortex core (within roughly twenty core radii), so escape dominates throughout most of the flow.
In the phase space, the fixed points at $\psi=\pi/2$ and $\psi=3\pi/2$ encode downstream and upstream rheotactic alignment, respectively. 
Escaping trajectories asymptotically approach the separatrix near $\psi=3\pi/2$, explaining how rheotactic bias and spatial depletion emerge simultaneously. 
Upstream migration occurs only once swimmers reach the boundary; during most of their escape, downstream alignment dominates. 
Rotational and translational diffusion intermittently transfer swimmers between trapping and escaping states, balancing upstream and downstream orientations while maintaining a net rheotactic preference.

Despite their different geometries, planar Poiseuille flow and rotlet flow admit a unified Smoluchowski description. In both cases, gradients of shear or vorticity generate a deterministic coupling between swimmer orientation and position, yielding an integrable two-dimensional phase space with conserved invariants. These invariants define separatrices that provide a deterministic explanation for depletion layers in Poiseuille flow and trapping in vortical flows.

\newpage
\section*{Supplementary Table 1. Parameters used in simulations and experiments}\label{ST1}
\begin{table*}[ht!]
	\centering
	\caption{\textbf{Simulation parameters used in each figure.} 
		Here, $W$ is the confinement width, $R$ the vortex-core radius, $V_s$ the swimming speed, 
		$D_t$ and $D_r$ the translational and rotational diffusivities, $B$ the alignment strength,
		$S$ the shear-alignment parameter, $\omega$ the angular velocity, and $f$ the rotation frequency.}
	\label{tab:simulation_parameters}
	\renewcommand{\arraystretch}{1.15}
	\begin{tabular}{lccccccccc}
		\hline\hline
		Figures &
		$W~(\um)$ &
		$R~(\um)$ &
		$V_s~(\um\,\text{s}^{-1})$ &
		$D_t$ &
		$D_r~(\text{rad}^2\,\text{s}^{-1})$ &
		$B$ &
		$S~(\text{rad}^{-1})$ &
		$\omega~(\text{rad}\,\text{s}^{-1})$ \\
		\hline
		Fig.~2a &
		425 & -- & 50 & 0 & 1 & 0.98 & -- & -- \\
		
		Fig.~2b &
		425 & -- & 50 & 0 & 1 & 0.98 & 10 & -- \\
		
		Fig.~3 &
		-- & 30 & 25 & -- & 0.1 & 1 & -- & $40\pi$ \\
		
		Fig.~4a,~b &
		425 & -- & 50 & 0 & 0 & 0.98 & 10 & -- \\
		
		Fig.~4c,~d &
		-- & -- & 0.1 & 0 & 0 & 0.5 & -- & 1 \\

		Ext.~Fig.~1a &
		425 & -- & 50 & 0 & 1 & 0.98 & -- & -- \\
		
		Ext.~Fig.~1b &
		-- & 30 & 25 & -- & 0.1 & 1 & -- & -- \\
		
		Ext.~Fig.~2 &
		-- & -- & 0.1 & 10 & 0 & 1 & -- & -- \\
		
		Ext.~Fig.~3 &
		-- & -- & 0.1 & -- & 0 & 1 & -- & 1 \\
		
		Ext.~Fig.~4 &
		-- & -- & 0.1 & 1 & 0 & 0 & -- & -- \\
		
		Ext.~Fig.~5 &
		-- & -- & 0.1 & 1 & 0 & 0 & -- & -- \\
		
		Ext.~Fig.~6 &
		-- & -- & 0.1 & 10 & 0 & 1 & -- & -- \\
		
		Suppl.~Fig.~1 &
		-- & -- & 0.1 & 10 & 0 & 1 & -- & -- \\
		
		Suppl.~Fig.~2a,~d &
		1 & -- & 1 & 0 & 0 & 0.98 & 1.1 & -- \\
		
		Suppl.~Fig.~2b,~e &
		425 & -- & 50 & 0 & 0 & 0.98 & 1.25 & -- \\
		
		Suppl.~Fig.~2c,~f &
		425 & -- & 50 & 0 & 0 & 0.98 & 10 & -- \\
		\hline\hline
	\end{tabular}
	
	\vspace{2mm}
	\raggedright
	Fig.~5 (experimental parameters):
	\textit{E.~coli} (wild type) $f=20$~Hz; 
	\textit{S.~pasteurii} $f=10$~Hz; 
	\textit{E.~coli} (smooth swimming) $f=20$~Hz.
\end{table*}

\newpage
\section*{Supplementary Videos 1--5}\label{SV}

\textbf{Supplementary Video 1 | Shear-induced depletion of \textit{E. coli} around a rotating particle (sessile drop).}
A rotating nickel particle (radius $R = 24.5,\mu\mathrm{m}$, rotation frequency 20 Hz) generates a three-dimensional vortical flow that reorients elongated \textit{Escherichia coli} cells and induces a systematic azimuthal drift. 
Particle rotation is driven by a uniform rotating magnetic field produced by two orthogonal pairs of Helmholtz coils. 
The resulting coupled orientation–position dynamics rapidly expel bacteria from the particle vicinity, forming a pronounced depletion zone. 
A small fraction of bacteria transiently accumulates near the particle surface; this near-surface accumulation is captured by a Robin boundary condition in the continuum model and disappears immediately when rotation ceases, confirming its hydrodynamic origin. 
Quantitative agreement is shown in \textbf{Extended Data Fig.}~3.\\

\textbf{Supplementary Video 2 | Processed visualization of Supplementary Video 1.}
The same experiment as in \textbf{Supplementary Video 1}, shown after background subtraction and noise reduction using ImageJ (\textbf{Supplementary Fig.}~3). 
Identical experimental parameters are used, with enhanced visibility of bacterial trajectories and depletion dynamics.\\

\textbf{Supplementary Video 3 | Shear-induced depletion of \textit{S. pasteurii} in a sessile drop geometry.}
A bacterial suspension (\textit{Sporosarcina pasteurii}, volume 10,$\mu$L) forms a sessile drop attached upon a microscope glass slide, providing a geometry that minimizes the influence of planar walls. 
A nickel particle (radius $R = 29.5,\mu\mathrm{m}$, rotation frequency 10,Hz) is positioned near the drop nadir by gravity. 
Despite variations in species-specific and morphological features, the vortical shear flow similarly reorients bacteria and expels them from the particle vicinity. 
The observed azimuthal drift is consistent with the unified theory, see \textbf{Extended Data Fig.}~3.\\

\textbf{Supplementary Video 4 | Processed visualization of Supplementary Video 3.}
The same experiment as in \textbf{Supplementary Video 3}, shown after background subtraction and noise reduction using ImageJ ((\textbf{Supplementary Fig.}~4).\\

\textbf{Supplementary Video 5 | Shear-induced depletion in a pendant drop geometry.}
A pendant drop of bacterial suspension (\textit{Bacillus subtilis}, volume 10,$\mu$L) is attached to a microscope glass slide. 
A rotating nickel particle (radius $R = 20,\mu\mathrm{m}$, rotation frequency 20,Hz) is held near the drop nadir by gravity. 
As in \textbf{Supplementary Videos 1--4}, the vortical shear flow induces bacterial reorientation and azimuthal drift, in agreement with Stokes-theory predictions without image correction.

\newpage
\section*{Supplementary Figure S1. Comparison of no-flow benchmark with simulations and continuum theory}

\begin{figure*}[ht!]
	\centering
	\includegraphics[width=0.9\linewidth]{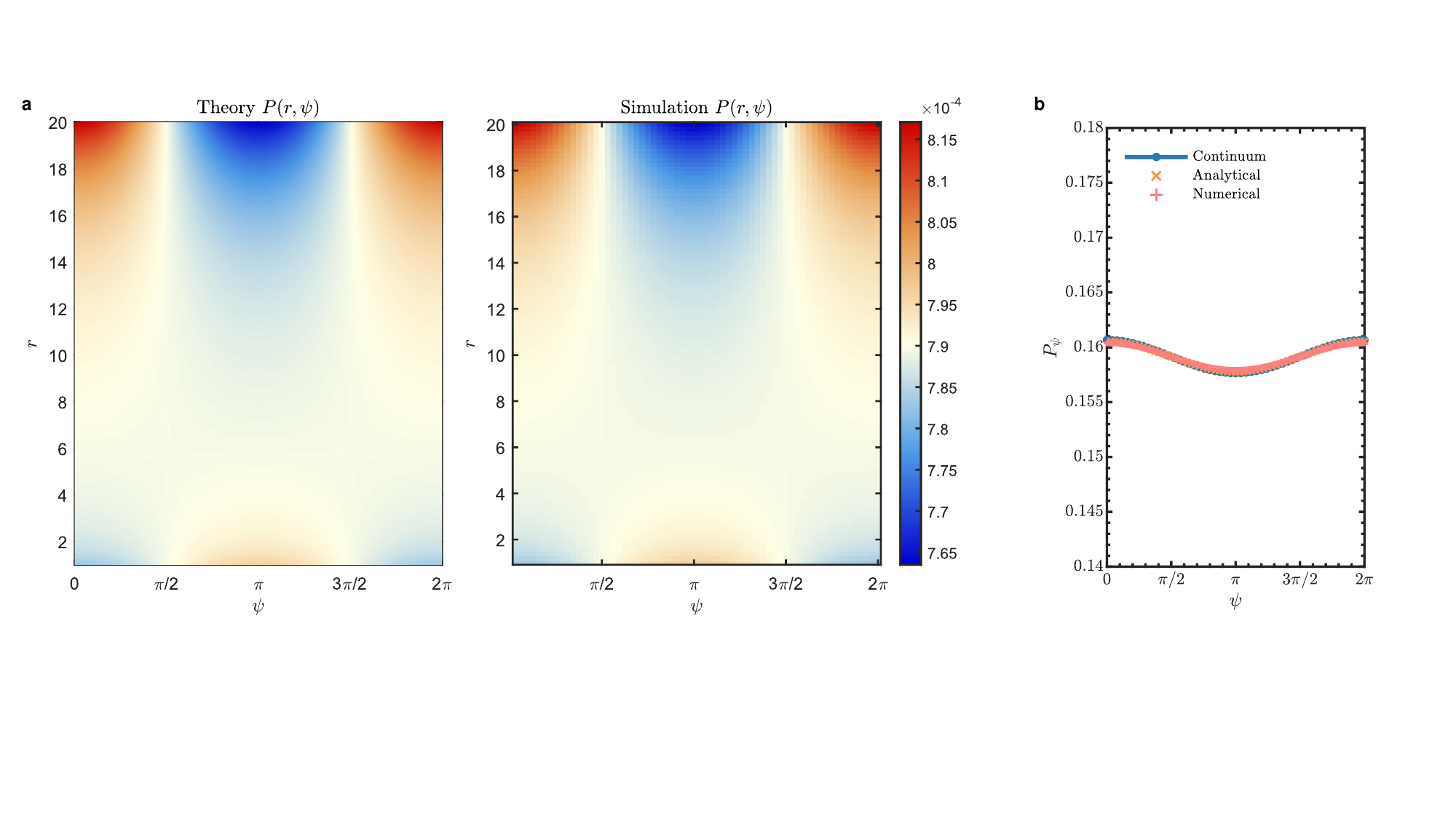}
	\caption{
		\textbf{ Steady-state distributions of bacteria in the no-flow case.}
		\textbf{a}, While \cref{eq_1d_sol_no_flow} may not depict the overall 2D distributions as the continuum theory does,
		the 1D analytical solution can well capture the average orientational distribution in \textbf{b}.
	}
	\label{figS1}
\end{figure*}

\newpage
\section*{Supplementary Figure S2. Shear compression of separatrices in various shear strengths}

\begin{figure*}[ht!]
	\includegraphics[width=0.9\linewidth]{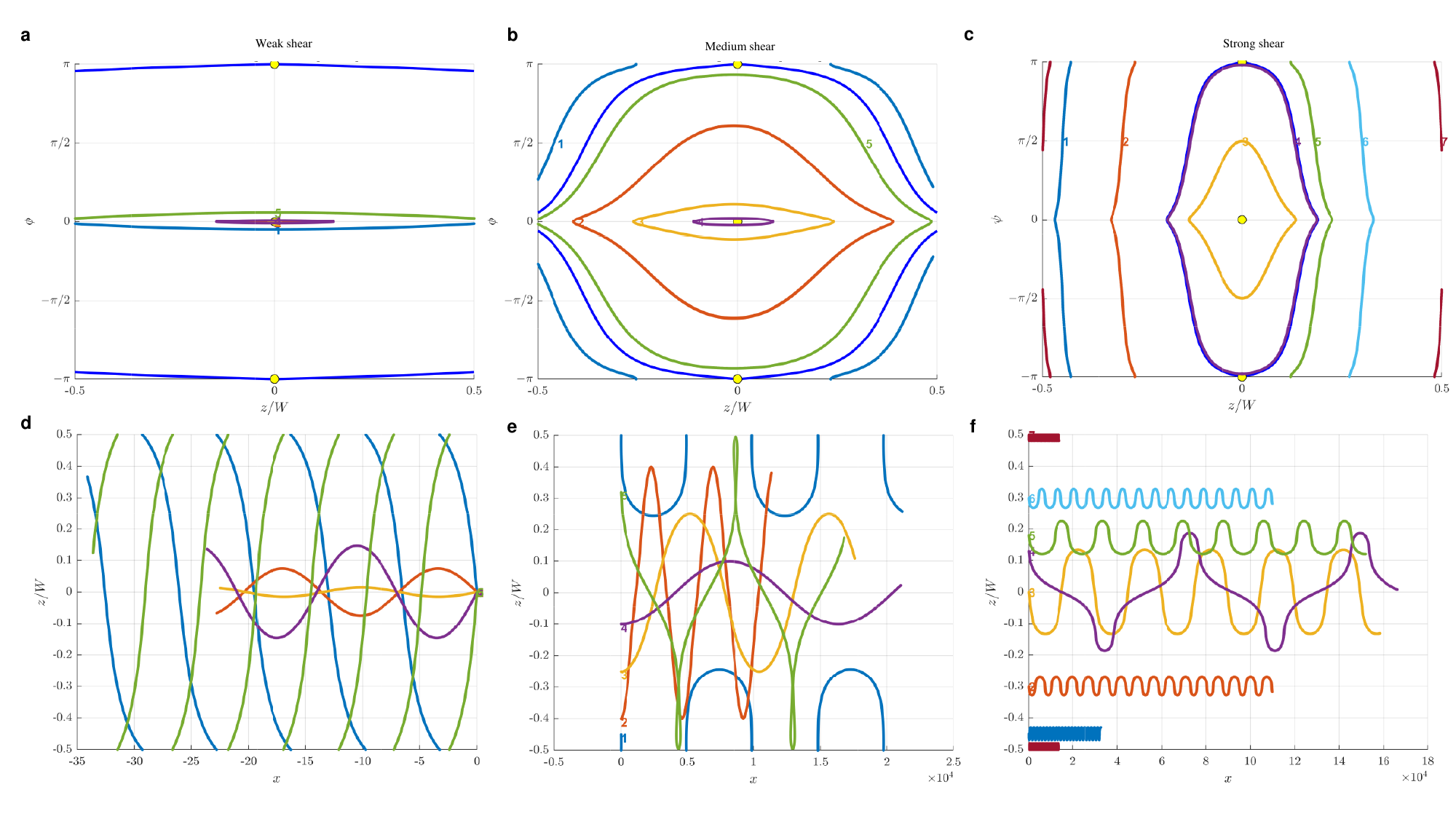}
	\caption{
		\textbf{Deterministic dynamics and separatrices of elongated micro-swimmers under various shear strengths.}
		Numbers indicate trajectories with different initial conditions and shear strengths.
		From left to right, the separatices are compressed into eye shapes as the mean shear rate increases.
		\label{figS2}}
\end{figure*}

\newpage
\section*{Supplementary Figure S3. Experimental observation of trapped \textit{E. coli}}

\begin{figure*}[ht!]
	\includegraphics[width=0.75\linewidth]{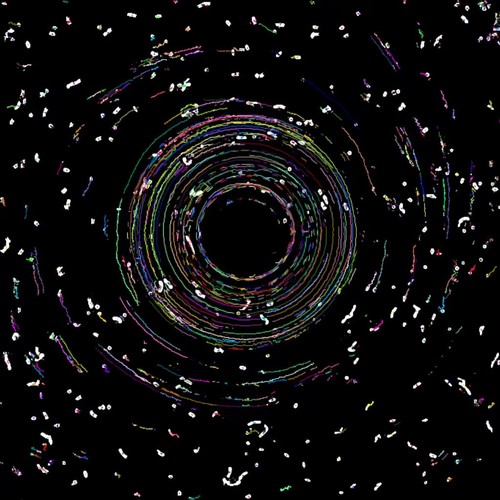}
	\caption{
		\textbf{Formation of hydrodynamic trapping of \textit{E. coli} with orientation locked to the background vortex.}
		Time-resolved trajectories reveal that bacteria align with and remain phase-locked to the background vortex, leading to persistent trapping and the emergence of a central depletion region.
		Images correspond to Supplementary Video 2 after background subtraction and noise reduction (ImageJ), with identical experimental parameters.
		\label{figS3}}
\end{figure*}

\newpage
\section*{Supplementary Figure S4. Reproducible vortex-driven trapping of \textit{S. pasteurii}}

\begin{figure*}[ht!]
	\includegraphics[width=0.75\linewidth]{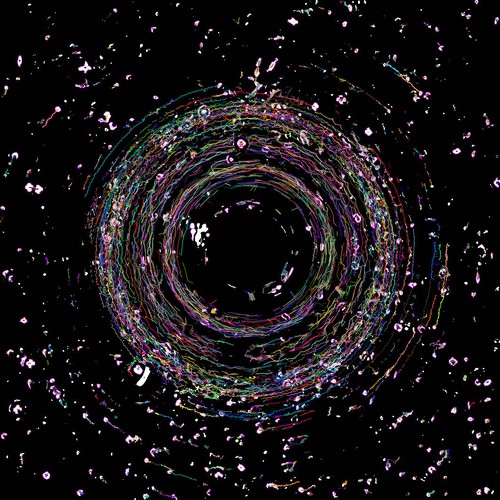}
	\caption{
		\textbf{Cross-species robustness of vortex-induced dynamics.}
		\textit{S. pasteurii} exhibits the same orientational locking and sustained hydrodynamic confinement observed in \textit{E. coli}, confirming that the depletion dynamics are not species-specific. 
		Images correspond to Supplementary Video 4 after identical post-processing.
		\label{figS4}}
\end{figure*}

\end{document}